\documentclass[superscriptaddress,secnumarabic,
amssymb,amsmath,nobibnotes,aps,prd,showkeys,showpacs,nofootinbib]{revtex4}%
\usepackage{graphicx}
\usepackage{epsf}
\usepackage{bm}
\usepackage{amsmath}
\usepackage{amsfonts}
\usepackage{amssymb}
\usepackage{epstopdf}
\usepackage{natbib}
\usepackage{color}
\providecommand{\U}[1]{\protect\rule{.1in}{.1in}}

\newcommand{\be}{\begin{equation}}
\newcommand{\ee}{\end{equation}}

\newcommand{\mincir}{\raise
-3.truept\hbox{\rlap{\hbox{$\sim$}}\raise4.truept\hbox{$<$}\ }}
\newcommand{\magcir}{\raise
-3.truept\hbox{\rlap{\hbox{$\sim$}}\raise4.truept\hbox{$>$}\ }}

\newtheorem{remark}{Remark}[section]

\begin{document}

\title{Understanding gravitational particle production in quintessential inflation}

\author{Jaume  de Haro}
\email{jaime.haro@upc.edu}
\affiliation{Departament de Matem\`atiques, Universitat Polit\`ecnica de Catalunya, Diagonal 647, 08028 Barcelona, Spain}

\author{Supriya Pan}
\email{supriya.maths@presiuniv.ac.in}
\affiliation{Department of Mathematics, Presidency University, 86/1 College Street, 
Kolkata 700073, 
India}

\author{Llibert Arest\'e Sal\'o}
\email{llibert.areste-salo@tum.de} 
\affiliation{Departament de Matem\`atiques, Universitat Polit\`ecnica de Catalunya, Diagonal 647, 08028 Barcelona, Spain}
\affiliation{TUM Physik-Department, Technische Universit{\"a}t M{\"u}nchen, James-Franck-Str.1, 85748 Garching, Germany}

\thispagestyle{empty}

\begin{abstract}
The diagonalization method, introduced by a group of  Russian scientists at the beginning of seventies,  is used to compute the energy density of superheavy massive particles produced due to a sudden phase transition from inflation to kination in quintessential inflation models, the models unifying inflation with quintessence originally proposed by Peebles-Vilenkin. These superheavy  particles must decay in lighter ones to form a relativistic plasma, whose energy density will eventually dominate the one of the inflaton field, in order to have a hot universe after inflation. In the present article we show that, in order that the overproduction of Gravitational Waves (GWs) during this phase transition does not disturb the Big Bang Nucleosynthesis (BBN) success,  the decay has to be produced after the end of the kination regime,
obtaining a { maximum} reheating temperature in the TeV regime.
\end{abstract}

\vspace{0.5cm}

\maketitle
\section{Introduction}

Understanding the universe's evolution has been a great mystery to modern cosmology. There are many questions related to different phases of the universe that are still undisclosed even after continuous investigations with different observational missions. 
In particular, its  early and late expansions have been a great deal at present time. Looking at the literature, one can find two popular and well accepted theories, namely the inflation (the early evolution of the universe) and the quintessence (the late evolution of the universe). The inflationary paradigm \cite{guth, linde,Burd:1988ss,Barrow:1990vx,Barrow:1994nt} is actually an accelerating phase of the early universe (in the context of standard Big Bang cosmology) that lasted for an extremely tiny time and  became able to solve a number of shortcomings associated with the standard Big Bang cosmology, such as the horizon problem, flatness problem and some more. The  potentiality of the inflationary theory was soon recognized due to its ability to explain the  origin of inhomogeneities in the universe \cite{chibisov, starobinsky, pi, bardeen, Linde:1982uu}. Such an explanation was found to match greatly with the  recent observational data from Planck \cite{Planck}. Thus, it is interesting to note that the theory that appeared at the beginning of the 80's is still surviving quite well with the recent observational data. And moreover, the theory of inflation is the simplest viable theory that describes almost correctly the early universe in agreement with the recent observations \cite{Planck}. On the other hand, the explanation for the current universe's expansion  comes through the introduction of some quintessence field \cite{Copeland:2006wr}. Thus, inflation  and quintessence were thought to be two different sides of a coin until the concept of the quintessential inflationary theory was introduced by Peebles and Vilenkin \cite{pv}.  

The idea to unify inflation with quintessence was indeed a novel attempt by Peebles and Vilenkin \cite{pv}. The novelty of their proposal comes through the introduction of a single potential that at early time allows inflation while at late time we have quintessence. Thus, a unified picture of the universe was effectively proposed connecting the distant early phase to the present one. 
{Thanks to this proposal, the origin of the scalar field responsible for the current {acceleration} of the universe can be determined, and the fine-tunning problems are reduced \cite{dimopoulos01}.
Moreover, as we will see, the models we deal with only depend on two real parameters, which are determined by the observational data. So, because of the behavior of the slow-roll regime as an attractor,  the dynamics of the model is obtained with the value of the scalar field, its derivative and the initial conditions at some moment during the inflation. This shows the simplicity of the quintessential inflation, which from our viewpoint is a little bit simplest than the standard quintessence, where a minimum of two fields are needed to depict the evolution of the universe, namely, the inflaton and a quintessence field, and thus, one needs two different potentials and two different initial conditions, one for the inflaton, which has to be fixed during inflation, and another for the quintessence field, whose initial conditions have to be fixed at the beginning of radiation era.  }

This enhanced more investigation in order to connect quintessential inflation with the observational data \cite{dimopoulos1,Giovannini:2003jw,hossain1,hossain3, deHaro:2016hpl,deHaro:2016hsh,deHaro:2016ftq,hap,Geng:2017mic,AresteSalo:2017lkv,Haro:2015ljc,hyp} and consequently this particular topic has become a popular area of research.    
The mechanism of the quintessential inflation model is very simple: once the inflationary phase is completed, a reheating mechanism is needed to match inflation with the hot Big Bang universe \cite{guth} because the particles existing before the beginning of this period were completely diluted at the end of inflation resulting in a very cold universe. 
The most accepted idea to reheat the universe in the context of quintessential inflation comes through an abrupt phase transition of the universe from inflation to kination (a regime where all the energy density of the inflation turns into kinetic \cite{Joyce}) where the adiabatic regime is broken and the particles are produced. The mechanism of  particle production is not unique in this context since a number of distinct mechanism are available and can be used.  The first one is the {\it gravitational particle production} studied long time ago  in \cite{Parker,fmm,glm,gmm,ford,Zeldovich}, at the end of the 90's in \cite{Damour, Giovannini} and more recently applied to quintessential inflation in \cite{Spokoiny, pv, dimopoulos0, vardayan} for massless particles. A second well-known  mechanism is the so-called {\it instant preheating} introduced in \cite{fkl0} and applied for the first time to  inflation in \cite{fkl}  and recently in \cite{dimopoulos, vardayan} in the context of $\alpha$-attractors in supergravity.  Other less popular mechanisms  are the {\it curvaton reheating} applied  to quintessence inflation in \cite{FL, ABM},
production of massive particles self-interacting and coupled to gravity \cite{tommi}
and the reheating via production of heavy massive particles conformally coupled to gravity \cite{kolb, kolb1,Birrell1, hashiba, hyp}. The production of superheavy massive particles is the primary concern of this  work. 
{ Our main motivation for using a conformally-coupled scalar field is its simplicity. Alternatively one could use other massive fields but the calculations would be more cumbersome. For instance, the Wentzel-Kramers-Brilloui (WKB) solution \footnote{We devote a full section on the use of WKB approximation} in equation 2.20 in \cite{Bunch}  gets considerably simplified when the scalar field is conformally coupled, i.e. $\xi=1/6$.}

In the Peebles-Vilenkin model \cite{pv}, the inflationary part is described by a quartic potential and, according to the recent observations, this does not suit well. To be explicit, for the quartic potential in the inflationary part of this potential the two-dimensional contour of ($n_s$, $r$) where $n_s$ is the 
scalar spectral index and $r$ is the ratio of tensor to scalar perturbations, 
does not enter into the 95\% confidence-level of Planck results \cite{Planck}. However, a simple change in the inflationary piece $-$ quartic to quadratic $-$ can solve this issue (see \cite{hap} for a detailed discussion and also see \cite{hyp}). On the other hand, 
the reheating mechanism followed in \cite{pv} is 
gravitational production of massless particles that results in a reheating temperature of  the  order  of  $1$ TeV. This reheating temperature is not
sufficient to solve the overproduction of the  Gravitational Waves (GWs). As a result the Big Bang Nucleosynthesis process can be hampered. 
Now, a lower bound for the reheating temperature comes in the following way. Since the
radiation-dominated  era occurs before the Big Bang  Nucleosynthesis (BBN) epoch which takes place in the  $1$ MeV  regime \cite{gkr}, the reheating temperature should naturally be greater than $1$ MeV.  But the upper bound of this reheating temperature is 
dependent on the theory we are concerned with. That means,  in some supergravity and superstring theories containing particles (for instance the gravitino or a modulus field) with only gravitational interactions, the thermal production of these relics and  its late time decay  may jeopardize the success of the standard BBN \cite{lindley}.
However, this problem can be avoided with the consideration of sufficiently low reheating temperature (of the order of $10^9$ GeV) \cite{eln}. 
Finally, one also needs to take into account that a viable reheating mechanism should deal with the pretension of the Gravitational Waves (GWs) in the BBN success that must satisfy the observational bounds appearing from the overproduction of the gravitational waves \cite{pv}.

Here we also consider a pre-heating due to the gravitational production of superheavy particles at the beginning of kination, where the inflationary and quintessence  pieces of the quintessential potential are matched. The heavy massive particles due to this pre-heating will start decaying in lighter ones to form a thermal relativistic plasma. We use the well-known Hamiltonian diagonalization method (see \cite{gmmbook} for a review) to calculate the energy density of the produced particles, showing that before the beginning of kination the vacuum polarization effects, which are geometric objects
associated to the creation and annihilation of the so-called {\it quasiparticles} \cite{gmmbook},  are sub-dominant and have no relevant effect in the Friedmann equation. On the contrary, after the abrupt phase transition to kination heavy massive particles are produced
and, since their energy density decreases as $a^{-3}$ before decaying in lighter particles and as $a^{-4}$ after that,
they will eventually dominate the energy density of the inflation whose decrease is as $a^{-6}$, and thus the universe will become reheated. 
Finally, we show that in our model the overproduction of GWs is compatible with the BBN success only 
when the decay of the superheavy  particles is after the end of the kination phase, leading to a reheating temperature of a few TeVs.

As usual we note that in the present manuscript we have worked on the units where $\hbar=c=1$ and the reduced Planck's mass is $M_{pl}\equiv \frac{1}{\sqrt{8\pi G}}\cong 2.4\times 10^{18}$ GeV.

\section{Creation of superheavy particles conformally coupled to gravity}
\label{sec-particle-creation}

\begin{figure}[ht]
\includegraphics[width=0.4\textwidth]{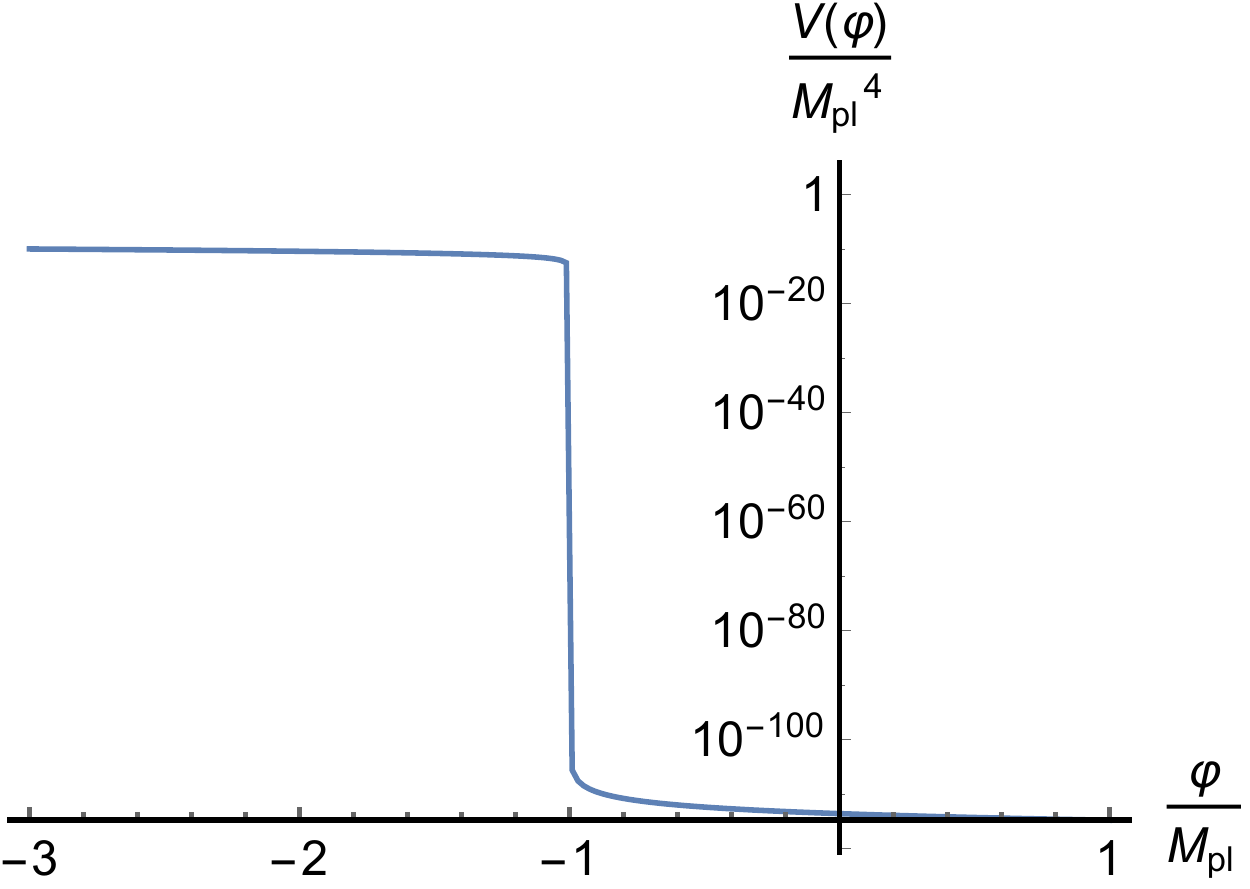}
\includegraphics[width=0.4\textwidth]{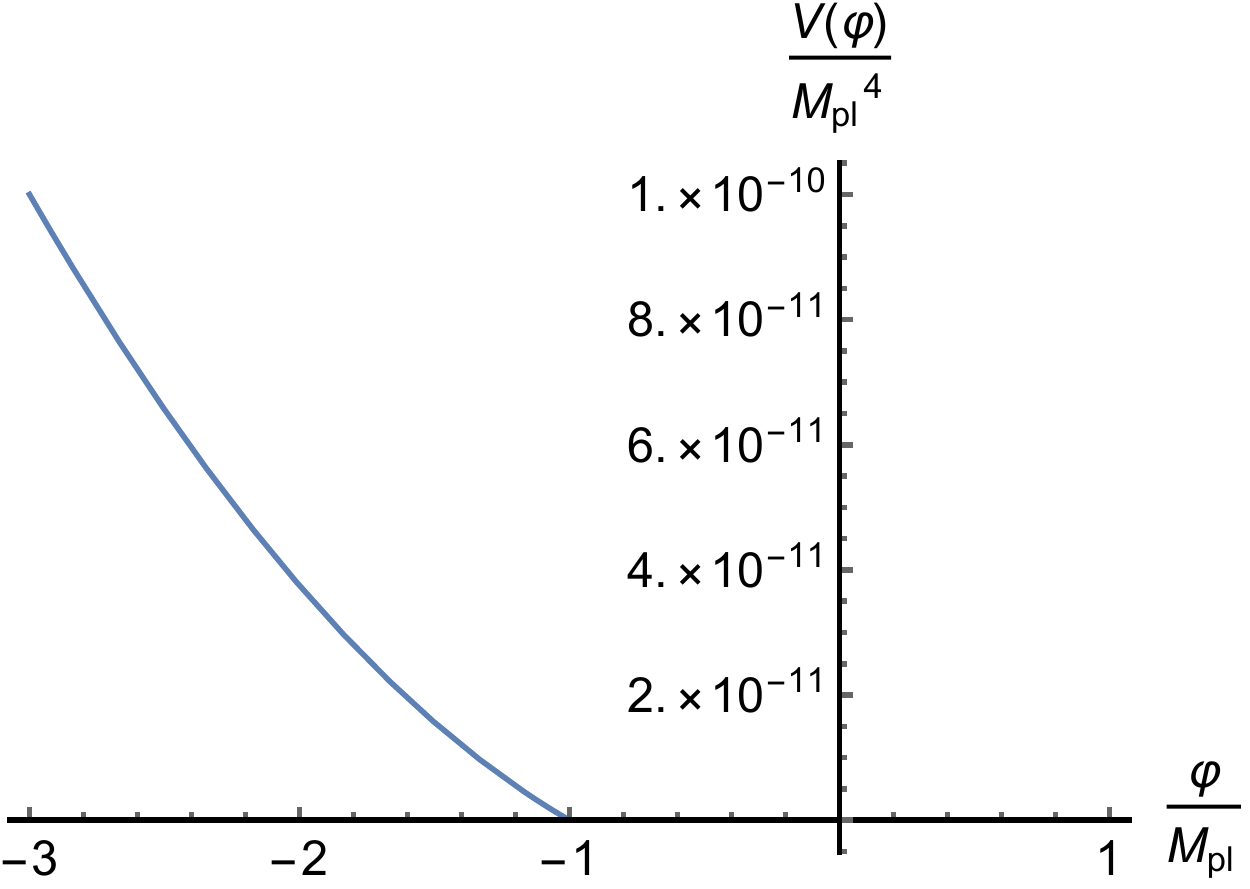}
\caption{{ The figure depicts the evolution of the first improved version of the Peeble-Vilenkin potential of eqn. (\ref{PV}), in two different scales, using the values derived in this section. }}
\label{fig:PV}
\end{figure}

\begin{figure}[ht]
\includegraphics[width=0.4\textwidth]{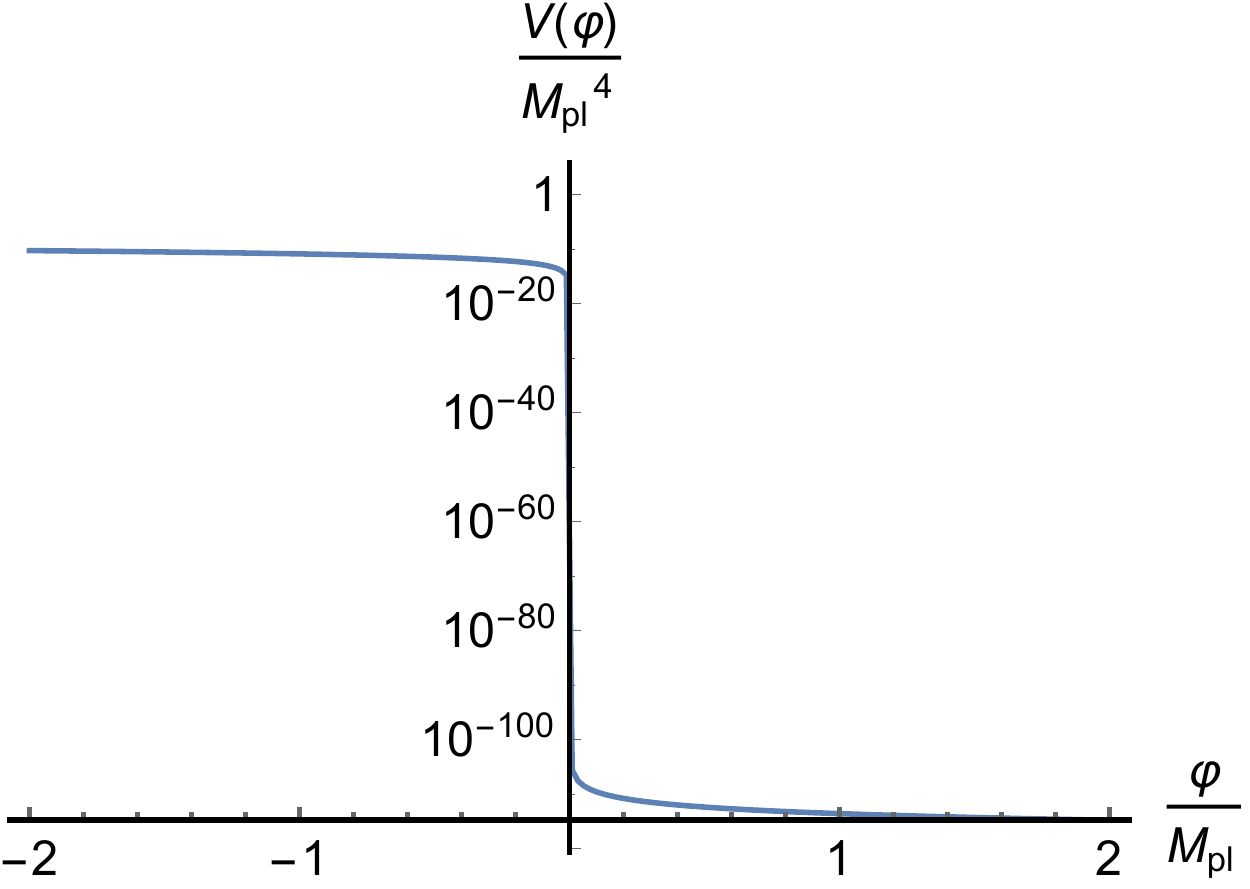}
\includegraphics[width=0.4\textwidth]{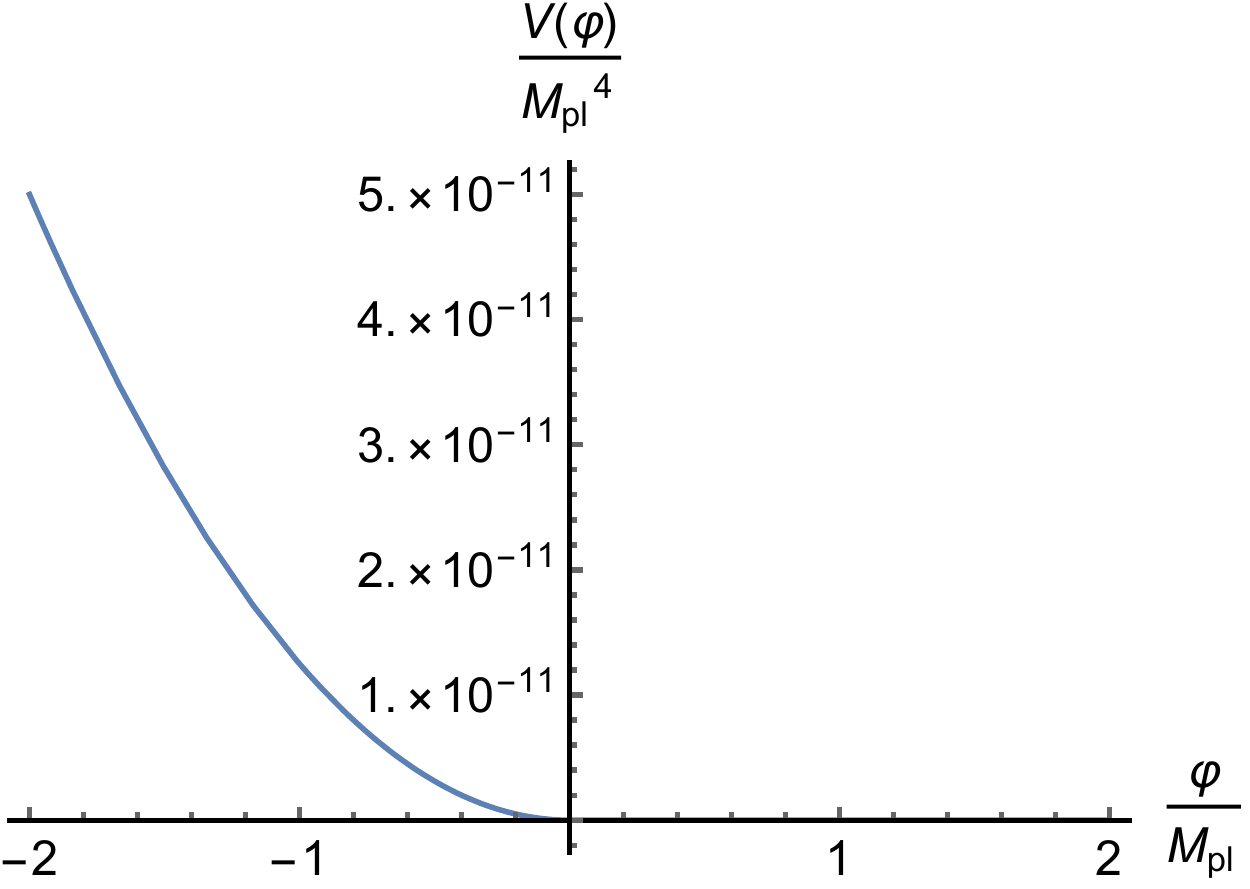}
\caption{{ The figure depicts the evolution of the first improved version of the Peeble-Vilenkin potential of eqn. (\ref{PV2}), in two different scales, using the values derived in this section. }}
\label{fig:PV2}
\end{figure}

{ In this section we shall describe the superheavy particles creation conformally coupled to gravity. Before that we refer to Appendix \ref{sec-diagonalization} (diaginalization method) and Appendix \ref{sec-wkb} (WKB approximation and its use in particle creation) which will be used throughout this work extensively. }
We begin this section with  the consideration of the models belonging  to the category of quintessential inflation with an abrupt phase transition from the end of inflation to the beginning of kination, as exactly in the Peebles-Vilenkin model \cite{pv}, where some of the higher order derivatives of $\omega_k(\tau)$ are discontinuous, which is essential for an efficient  production of 
superheavy particles. { Otherwise, if we had a smooth transition, the production of particles would be exponentially suppressed \cite{kolb} and
its energy density would be abnormally small. Therefore, it would never dominate those of the background, which means that the universe would never be reheated}. 
In this way, the  two quintessential inflationary models considered in this work are the improvements of the well known Peebles-Vilenkin model as follows:
\begin{enumerate} 
\item The first quintessential inflationary model that we consider is,
\begin{eqnarray}\label{PV}
V(\varphi)=\left\{\begin{array}{ccc}
\frac{1}{2}m^2\left(\varphi^2-M_{pl}^2+M^2\right)& \mbox{for}& \varphi\leq -M_{pl},\\
\frac{1}{2}m^2\frac{M^6}{(\varphi+M_{pl})^4+M^4}& \mbox{for}& \varphi\geq -M_{pl}.
\end{array} \right.
\end{eqnarray}
\item The second quintessential inflationary model in this work is,
\begin{eqnarray}\label{PV2}
V(\varphi)=\left\{\begin{array}{ccc}
\frac{1}{2}m^2(\varphi^2+M^2)& \mbox{for}& \varphi\leq 0\\
\frac{1}{2}m^2\frac{M^6}{\varphi^4+M^4} &\mbox{for}& \varphi\geq 0.
\end{array}\right. 
\end{eqnarray}
\end{enumerate}
{ While to understand the behavior of the above two modified potentials, we plot them in Fig. \ref{fig:PV} [for eqn. (\ref{PV})] and Fig. \ref{fig:PV2} [for eqn. (\ref{PV2})] in two different scales in order to exactly show the abrupt phase transition. The left panels of both Fig. \ref{fig:PV} and Fig. \ref{fig:PV2} are drawn in higher scale while the right panels of Fig. \ref{fig:PV} and Fig. \ref{fig:PV2} are for smaller scales. Let us note that while drawing the plots we have used the derived values of other parameters, namely, $m$ and $M$, shown in the latter part of this section. }

The  inflation's mass $m$ is obtained from the power spectrum of the curvature fluctuation in co-moving coordinates when the pivot scale leaves the Hubble radius \cite{btw}, given by
${\mathcal P}_{\zeta}\cong \frac{H_*^2}{8\pi^2 M_{pl}^2\epsilon_*}\sim 2\times 10^{-9}$, where $\epsilon=\frac{M_{pl}^2}{2}\left(\frac{V_{\varphi}}{V}\right)^2$ is a slow roll parameter and the ``star'' ($\ast$) attached to any quantity means that the quantity is evaluated when the pivot scale leaves the Hubble radius. For the first potential 
one has $\epsilon_* =  \frac{2M_{pl}^2\varphi_*^2}{(\varphi_*^2-M_{pl}^2)^2}    \cong\frac{2M_{pl}^2}{\varphi^2_*}$, where we have used that
$-\varphi_*\gg M_{pl}$. In the same way 
$\eta_*=M_{pl}^2\frac{V_{\varphi\varphi}}{V}\cong\frac{2M_{pl}^2}{\varphi^2_*}$, and since the spectral index is given by
$1-n_s=6\epsilon_*-2\eta_*$ one gets $\epsilon_*\cong \frac{1-n_s}{4}$. Finally, since at the time of the inflation the energy density is dominated by the potential term, using the Friedmann equation  
$H_*^2={\frac{V(\varphi_*)}{3M_{pl}^2}}$ one has
\begin{eqnarray}
m^2\sim 3\times 10^{-9} \pi^2(1-n_s)^2   M_{pl}^2.
\end{eqnarray}

\begin{figure}
\includegraphics[width=0.6\textwidth]{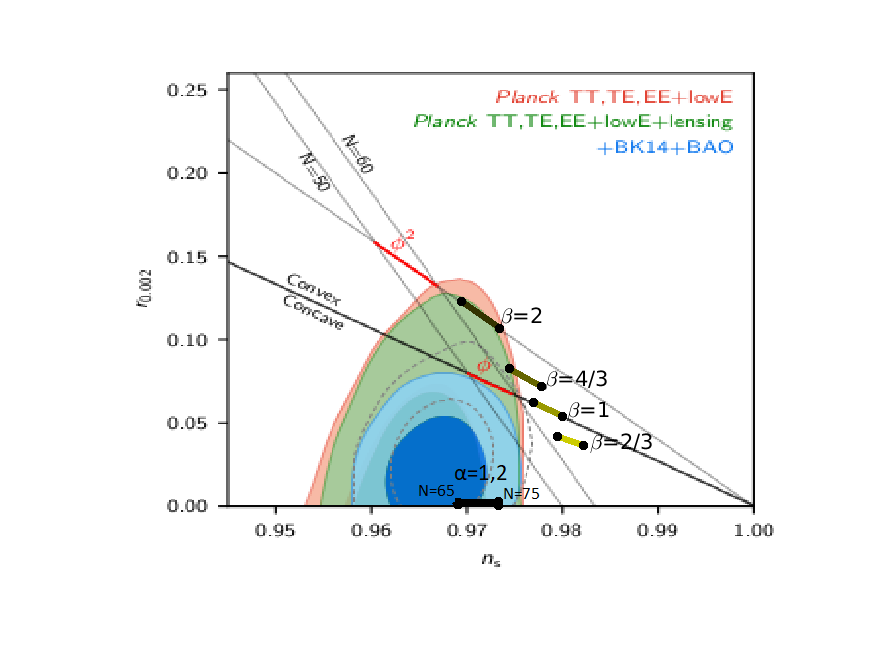}
\caption{{ Marginalized joint confidence contours for $(n_s,r)$ at $68\%$ and $95\%$ confidence level.  Considering the inflationary piece of the potential as 
$V=\lambda \phi^{\beta}$, in
quintessential inflation, for the values of $\beta=2, 3/4, 1, 2/3$, we have
drawn the curves  from  $65$ to $75$ e-folds (see the black curves). And  when one considers the standard inflation, for $\beta=2, 1$, the curves have been drawn in red from $50$ to $60$ $e$-folds.
As one can see, the quadratic potential ($V \propto \phi^2$), which is disregarded in standard inflation at greater than 95\% CL
from a combination of Planck and BICEP2 limits on the tensor-to-scalar 
ratio \cite{Akrami:2018odb}, is favored  for some likelihoods in quintessential inflation. In the lower part of the image there are the curves for the values of $\alpha=1,2$ for the potentials \eqref{susy} and \eqref{higgs}. The value of $r$ is nearly $0$ and, if considering all Planck likelihoods, they stand within the $1\sigma$ CL for a low number of e-folds ($65\lesssim N\lesssim 70$), while they are in the $2\sigma$ CL for the other values of $N$.}}
\label{fig:n-r}
\end{figure}

 Thus, since recent observations constrain the value of the spectral index to be $n_s=0.968\pm 0.006$ \cite{Planck}, hence, taking its central value  one can evaluate $m\cong 5\times 10^{-6} M_{pl}$.  The other parameter $M$ is a very small mass compared to the reduced Planck's mass $M_{pl}$, whose numerical value is determined so that 
at the present time the ratio of the energy density of the inflaton field $\varphi$ to the critical energy density is approximately around $0.7$ \cite{Planck}, that means, 
$\rho_{\varphi,0}/(3H_0^2M_{pl}^2) \cong 0.7$, where the sub-index $0$ means ``at present time'' and 
$\rho_{\varphi}=\dot{\varphi}^2/2+V(\varphi)$ is the energy density of the inflaton field. Numerical calculations performed in  \cite{hap18}  show that the value of $M$ depends on the reheating temperature
and for a reheating temperature of the order of $100$ TeV, which is the one obtained when the reheating is due to the production of superheavy  particles \cite{hyp}, one gets 
$M\sim 18$ GeV. 
{ Moreover, as  we show in Figure \ref{fig:n-r}
the values of the power spectrum and the ratio of tensor to scalar perturbations stand within $2\sigma$ confidence level for some given Planck likelihoods but not if we consider all the ones available in the 2018 Planck results \cite{planck18}. For that purpose, one would need to consider plateau potentials \cite{plateau} or $\alpha$-attractors \cite{attractor1,attractor2}, such as an Exponential SUSY Inflation type potential}
{
\begin{eqnarray} \label{susy}
V_{\alpha}(\varphi)=\left\{\begin{array}{ccc}
\lambda M_{pl}^4\left( 1-e^{\alpha\varphi/M_{pl}} + \left(\frac{M}{M_{pl}}\right)^4\right) & \mbox{for} & \varphi\leq 0\\
\lambda\frac{M^8}{\varphi^{4}+M^4} &\mbox{for} & \varphi\geq 0,\end{array}
\right.
\end{eqnarray}}
{ or, a Higgs Inflation-type potential}{
\begin{eqnarray} \label{higgs}
V_{\alpha}(\varphi)=\left\{\begin{array}{ccc}
\lambda M_{pl}^4\left( 1-e^{\alpha\varphi/M_{pl}} + \left(\frac{M}{M_{pl}}\right)^2\right)^2 & \mbox{for} & \varphi\leq 0\\
\lambda\frac{M^8}{\varphi^{4}+M^4} &\mbox{for} & \varphi\geq 0.\end{array}
\right.
\end{eqnarray}}
{ For both potentials one can calculate the spectral index and the ratio of tensor to scalar perturbations, obtaining}
{
\begin{eqnarray}
n_s\cong 1-\frac{2}{N},\qquad r\cong \frac{8}{\alpha^2 N^2},
\end{eqnarray}}
{ which implies that for $\alpha\sim {\mathcal O}(1)$ and for a number of $e$-folds greater than $60$, which is typic in quintessential inflation due to the kination phase,  the ratio of tensor to scalar perturbations is less than $0.003$. Thus, the spectral index and the tensor/scalar ratio 
enter perfectly in the  two dimensional marginalized joint confidence contour at $2\sigma$ CL for the Planck TT, TE, EE + low E + lensing + BK14 + BAO likelihood.}

\ 

The dynamics of the first potential (and also the second one) is not difficult to understand.  When $\varphi\ll -M_{pl}$, the field slowly rolls and thus the universe inflates; after the inflation
a phase transition  from inflation to kination \cite{Joyce}
occurs about  $\varphi\cong -M_{pl}$ and the particles are produced.
Since in a kination  regime  the energy density of the background  decays as $a^{-6}$, this  allows a relativistic plasma in thermal equilibrium, whose energy density evolves as $a^{-4}$, to eventually become dominant, and the universe is thus reheated. Finally, at the present time, the potential energy  of the scalar field $\varphi$ becomes dominant once again and the universe accelerates, depicting the current cosmic acceleration. Thus, as a result we have a unified framework where at the early time the universe experiences a rapid accelerating phase and at late time another accelerating phase leading to the current dark energy era. 
Note also that for the second model the second derivative of the potential is discontinuous at the beginning of kination. So, using the conservation equation, one can deduce that the third temporal derivative of the scalar field is discontinuous at the begininning of kination, as well as the third temporal derivative of the Hubble parameter, as one can infere from Raychaudhuri equation. The first potential is more abrupt and at the beginning of kination the second derivative of the Hubble parameter is discontinuous.  So, dealing with the first
one, the third derivative of the frequency $\omega_k(\tau)$ is discontinuous at the beginning of kination, namely $\tau_{kin}$. 

\
  
A key point is related to the initial conditions. It is well-known that at temperatures of the  order of the Planck's mass quantum effects become very important  and the classical picture of the universe is not possible of course. However, at temperatures below $M_{pl}$, for example 
at GUT scales (i.e., when the temperature of the universe is of the order of $T\sim 4\times 10^{-3} M_{pl}\sim  10^{16}$ GeV), the beginning of the  Hot Big Bang (HBB) scenario is possible. Since for the  flat FLRW universe the energy density of the universe, namely $\rho$,  and the Hubble parameter $H$ of the FLRW universe are related through the Friedmann equation $\rho=3H^2M_{pl}^2$ and the temperature of the universe is related to the energy density via $\rho = (\pi^2/30)g_{*} T^4$ (where  $g_*=106.75$ is the number degrees of freedom for the energy density in the Standard Model), one can conclude that a classical picture of the universe might be possible when $H\cong 5\times 10^{-5} M_{pl}\cong 10^{14}$ GeV. Then, 
if inflation starts at this scale, i.e. taking the value of the Hubble parameter at the beginning of inflation as $H_{B}=5\times 10^{-5} M_{pl}$,
we will assume as a natural initial condition that {a superheavy massive quantum $\chi$-field, whose decay products are the responsible of 
the reheating of the universe},  is in the vacuum at the beginning of inflation.
We will also choose  the mass of the  $\chi$-field one order greater than this value of  the Hubble parameter ($m_{\chi}=5\times 10^{-4} M_{pl}\cong 10^{15}$ GeV, which is a mass of the same order as those of the vector mesons responsible to transform quarks into leptons in simple theories with  SU(5)  symmetry \cite{lindebook}) because, as we will immediately see, the polarization terms will be sub-dominant and do not affect the dynamics of the inflation field.
So, we have $m\ll H_{B}\ll m_{\chi}\ll M_{pl}$.

\

{ To obtain the energy density of the produced particles by the $\chi$-field (see formula (A9) of Appendix A) we have to calculate the
value of the $\beta$-Bogoliubov coefficient, whose expression has been derived in formula   (A10). To perform it, } we have to integrate by parts two times, then before the beginning of kination one has 
\begin{eqnarray}
\beta_k(\tau)=-\frac{\omega'_k(\tau)}{4i\omega_k^2(\tau)}e^{-2i\int^{\tau} \omega_k(\bar\eta)d\bar\eta}+
\int^{\tau}\left(\frac{\omega'_k(\eta)}{4i\omega_k^2(\eta)}\right)'e^{-2i\int^{\eta} \omega_k(\bar\eta)d\bar\eta}d\eta\nonumber \\
= \left(-\frac{\omega'_k(\tau)}{4i\omega_k^2(\tau)}+\frac{1}{8\omega_k(\tau)}\left(\frac{\omega'_k(\tau)}{\omega_k^2(\tau)}\right)'
+\frac{1}{16i\omega_k(\tau)}\left(\frac{1}{\omega_k(\tau)}\left(\frac{\omega'_k(\tau)}{\omega_k^2(\tau)}\right)'\right)'+....
  \right)e^{-2i\int^{\tau} \omega_k(\bar\eta)d\bar\eta}.
  \end{eqnarray}

However, after kination the $\beta$-Bogoliubov coefficient, in order to be continuous in time must be given by
\begin{eqnarray}\label{Bogoliubov}
\beta_k(\tau)=
\left(-\frac{\omega'_k(\tau)}{4i\omega_k^2(\tau)}+\frac{1}{8\omega_k(\tau)}\left(\frac{\omega'_k(\tau)}{\omega_k^2(\tau)}\right)'
+\frac{1}{16i\omega_k(\tau)}\left(\frac{1}{\omega_k(\tau)}\left(\frac{\omega'_k(\tau)}{\omega_k^2(\tau)}\right)'\right)'+....
  \right)e^{-2i\int^{\tau} \omega_k(\bar\eta)d\bar\eta}+C,\end{eqnarray}
where the constant $C$ has to be chosen in order that the $\beta$-Bogoliubov coefficient becomes continuous at $\tau_{kin}$ because the equation (\ref{Bogoliubovequation}) [see Appendix \ref{sec-diagonalization}]
is a first order differential equation. So, one has to impose continuity at the beginning of kination in the same way that happens when one matches the modes. In this case, since they satisfy the second order K-G differential equation, the matching involves the continuity of the first derivative.  Therefore, for  the first potential one has
\begin{eqnarray}\label{constant}
C=
\left(
\frac{1}{16i\omega_k(\tau^-_{kin})}\left(\frac{1}{\omega_k(\tau^-_{kin})}\left(\frac{\omega'_k(\tau^-_{kin})}{\omega_k^2(\tau^-_{kin})}\right)'\right)'-
\frac{1}{16i\omega_k(\tau^+_{kin})}\left(\frac{1}{\omega_k(\tau^+_{kin})}\left(\frac{\omega'_k(\tau^+_{kin})}{\omega_k^2(\tau^+_{kin})}\right)'\right)' +....
  \right)e^{-2i\int^{\tau_{kin}} \omega_k(\bar\eta)d\bar\eta}\nonumber \\
  =\left(
\frac{\omega_k'''(\tau_{kin}^-)-\omega_k'''(\tau_{kin}^+)}{16i\omega^4_k(\tau_{kin})}+
....
  \right)e^{-2i\int^{\tau_{kin}} \omega_k(\bar\eta)d\bar\eta}=
  \left(
\frac{m_{\chi}^2a_{kin}(a'''(\tau_{kin}^-)-a'''(\tau_{kin}^+))}{16i\omega^5_k(\tau_{kin})}+
....
  \right)e^{-2i\int^{\tau_{kin}} \omega_k(\bar\eta)d\bar\eta} \nonumber \\
  =
   \left(
\frac{m_{\chi}^2a^5_{kin}(\ddot{H}(\tau_{kin}^-)-\ddot{H}(\tau_{kin}^+))}{16i\omega^5_k(\tau_{kin})}+
....
  \right)e^{-2i\int^{\tau_{kin}} \omega_k(\bar\eta)d\bar\eta}
  =
   \left(
\frac{m_{\chi}^2m^3a^5_{kin}}{16i\omega^5_k(\tau_{kin})}+
....
  \right)e^{-2i\int^{\tau_{kin}} \omega_k(\bar\eta)d\bar\eta},
 \end{eqnarray}
where $a_{kin}\equiv a(\tau_{kin})$ and having used that
\begin{eqnarray}
\ddot{H}(\tau_{kin}^+)-\ddot{H}(\tau_{kin}^-)=-\frac{\dot{\varphi}_{kin}}{M_{pl}^2}(\ddot{\varphi}(\tau_{kin}^+)- \ddot{\varphi}(\tau_{kin}^-))
=-\frac{\dot{\varphi}_{kin}}{M_{pl}^2}V_{\varphi}(-M_{pl}^-)= \frac{m^2\dot{\varphi}_{kin}}{M_{pl}}=
m^3, \quad \mbox{with}  \quad \dot{\varphi}(\tau_{kin})\equiv \dot{\varphi}_{kin}, 
 \end{eqnarray}
with the assumption that  there is no substantial drop of energy density between the end of inflation and the beginning of kination. 
Thus, at $\tau_{kin}$ all the energy density is kinetic and given by $\frac{1+\sqrt{3}}{2}m^2 M_{pl}^2$ because at the end of inflation, where all the energy density is potential, one has  $\varphi_{end}=-\sqrt{2+\sqrt{3}}M_{pl}$.
The terms that do not contain $C$ lead to a sub-leading geometric quantities in the energy density. Effectively, the term $-\frac{\omega'_k(\tau)}{4i\omega_k^2(\tau)}$
leads to the following contribution to the energy density $\frac{m_{\chi}^2 H^2}{96 \pi} \ll 3M_{pl}^2H^2$. The same happens with the term $\frac{1}{8\omega_k(\tau)}\left(\omega'_k(\tau)/\omega_k^2(\tau)\right)'$ which leads to a term of order $H^4$, which means that $\frac{H^4}{M_{pl}^2}\ll H^2$.
The product of the first and second term generates in the right-hand side of the modified semi-classical Friedmann  equation a term of the order 
$\frac{H^3m_{\chi}}{M_{pl}^2}$, which is also sub-leading compared with $H^2$. Finally, the third term of (\ref{Bogoliubov}) leads in the right-hand side of the semi-classical Friedmann equation to the sub-leading term $\frac{H^6}{m_{\chi}^2M_{pl}^2}$.

\

Fortunately, this does not happen with $C$, whose leading term is $\frac{m_{\chi}^2m^3a^5_{kin}}{16i\omega^5_k(\tau_{kin})}$, leading to the contribution  (see \cite{hyp} 
and the appendix of \cite{ha} for a detailed derivation of this result)
\begin{eqnarray}
\langle\rho(\tau)\rangle\cong \left\{\begin{array}{ccc}
0& \mbox{ when} & \tau< \tau_{kin} \\
10^{-5}\left(\frac{m}{m_{\chi}}  \right)^2m^4\left( \frac{a_{kin}}{a(\tau)} \right)^3 & \mbox{ when}    & \tau\geq \tau_{kin},
\end{array}\right.
\end{eqnarray}
which at the beginning of kination is sub-dominant with respect to the energy density of the background but eventually it will dominate because the one of the background, during kination, 
decreases as $a^{-6}(\tau)$.

\begin{remark}
The authors of the diagonalization method assume that during the whole evolution of the universe quanta named {\it quasiparticles} are created and annihilated due to the interaction with the quantum field with gravity \cite{gmmbook}. Following this interpretation, the number density of  created {\it quasiparticles} at time $\tau$ is given by
$\langle N(\tau)\rangle=\frac{1}{2\pi^2 a^3(\tau)}\int_0^{\infty}k^2 |\beta_k(\tau)|^2 dk$. However, one has to be very careful with this interpretation and specially keep in mind that
real particles are only created when the adiabatic regime breaks. Effectively, before the beginning of kination the main term of the $\beta_k$-Bogoliubov coefficient is given by $-\frac{\omega'_k(\tau)}{4i\omega_k^2(\tau)}$,
whose contribution to the energy density is $\frac{m_{\chi}^2 H^2}{96 \pi}$, and to the number density of {\it quasiparticles} is $\frac{m_{\chi}^2 H^2}{512 \pi}$, and thus,
at time $\tau$ before the beginning of kination $\langle\rho(\tau)\rangle\not=m_{\chi}\langle N(\tau)\rangle$. On the contrary, during kination
the leading term of 
$\langle N(\tau)\rangle$ is given by $ 10^{-5}\left(\frac{m}{m_{\chi}}  \right)^3m^3\left( \frac{a_{kin}}{a(\tau)} \right)^3$, so we have $\langle\rho(\tau)\rangle=m_{\chi}\langle N(\tau)\rangle$
and the decay follows $a^{-3}(\tau)$,
which justifies the interpretation of massive particle production.
\end{remark}

Finally, for the second potential a similar calculation leads to 
 \begin{eqnarray}
 |\beta_k(\tau)|^2\cong 
 \frac{m_{\chi}^4a^{12}(\tau_{kin})(\dddot{H}(\tau_{kin}^-)-\dddot{H}(\tau_{kin}^+))^2}{1024\omega^{12}_k(\tau_{kin})}= 
 \frac{m_{\chi}^4m^8a^{12}_{kin}}{256\omega^{12}_k(\tau_{kin})},\end{eqnarray}
and a simple calculation shows that, { after the beginning of kination}, the energy density is given by
\begin{eqnarray}
\langle\rho(\tau)\rangle\cong 8\times10^{-6}\left(\frac{m}{m_{\chi}}  \right)^4m^4\left( \frac{a_{kin}}{a(\tau)} \right)^3,\end{eqnarray}
which is smaller than the one obtained from the first, more abrupt,  potential.

\section{The reheating process}
\label{sec-reheating}

After the production of the heavy massive particles, they have to decay in lighter particles which after the thermalization process form a relativistic plasma that depicts our hot universe. Two different situations may arise, as follows: 
\begin{enumerate}
\item The decay is before the end of the kination regime, which happens at time $\tau_r$, when the energy density of the inflaton becomes equal to the one of the $\chi$-field.
\item The decay is after the end of the kination regime.
\end{enumerate}

Here we consider the decay of the  $\chi$-field into fermions ($\chi\rightarrow \psi\bar\psi$), then the decay rate will be given by \cite{lindebook}
${\Gamma}=\frac{h^2 m_{\chi}}{8\pi}$ and the decay is finished at $\tau_{dec}$ when $\Gamma\sim H(\tau_{dec})\equiv H_{dec}$.

\subsection{Decay before the end of kination}
\label{sebsec1-reheating}

Let us begin the discussion with the first potential. In this case,
 the energy density of the background, i.e. the one of the inflaton field,  and the one of the relativistic plasma, when the decay is finished, 
that is
 when ${\Gamma}\sim H_{dec}=H_{kin}\left(\frac{a_{kin}}{a_{dec}} \right)^3\cong
  \frac{\sqrt{1+\sqrt{3}}}{\sqrt{6}}m\left(\frac{a_{kin}}{a_{dec}}\right)^3 $, will be
\begin{eqnarray}
\rho_{\varphi, dec}=3{\Gamma}^2M_{pl}^2 \quad \mbox{and} \quad \langle\rho_{dec}\rangle\cong 1.5\times 10^{-5}\left( \frac{m}{m_{\chi}} \right)^2 \frac{{\Gamma}}{m}m^4,
\end{eqnarray}
where we have used that there is no drop of energy density between the end of inflation and the beginning of kination, i.e., $H^2(\tau_{kin})\equiv H_{kin}^2=\frac{1+\sqrt{3}}{6}m^2$.

Imposing that the end of the decay precedes the end of kination, that means, $ \langle\rho_{dec}\rangle\leq \rho_{\varphi, dec}$, one gets
\begin{eqnarray}\label{BOUND}
h^2\geq {4\pi}\times 10^{-5}\left( \frac{m}{m_{\chi}} \right)^3 \left(\frac{m}{M_{pl}}  \right)^2,
\end{eqnarray}
which, for the value of the inflaton mass $m\cong 5\times 10^{-6}M_{pl}$ and  the bare mass of the quantum field 
 $m_{\chi}\cong 5\times 10^{-4}M_{pl}$,
constrains the
value of the coupling constant as $h\geq  5.6\times10^{-11}$. Moreover, since the decay is after the beginning of the kination, one has
${\Gamma}\leq H_{kin}$, obtaining $h^2\leq \frac{8\pi H_{kin}}{m_{\chi}}$, which for the values of $H_{kin}$ and $m_{\chi}$ gives another restriction, namely $ h\leq 3.8\times 10^{-1}$. Thus, we have obtained that the parameter $h$ is constrained as  $5.6\times 10^{-11}\leq h \leq 3.8\times 10^{-1}$.

Then the reheating temperature (i.e., the temperature of the universe when the relativistic plasma in thermal equilibrium starts to dominate, 
which happens when $\rho_{\varphi, reh}=\langle \rho_{reh}\rangle\Longleftrightarrow \frac{\langle \rho_{dec}\rangle}{\rho_{\varphi,dec}}=\left(a_{dec}/a_{reh}\right)^2$)
will be 
\begin{eqnarray}\label{reheating1}
 T_{reh}=  \left(\frac{30}{\pi^2g_*} \right)^{1/4}
 \langle\rho_{reh}\rangle^{\frac{1}{4}}= 
 \left(\frac{30}{\pi^2g_*} \right)^{1/4}
 \langle\rho_{dec}\rangle^{\frac{1}{4}}
 \sqrt{\frac{\langle\rho_{dec}\rangle}{\rho_{\varphi,dec}}} 
\cong {2\times 10^{-4}} g_*^{-1/4}\left(\frac{m}{m_{\chi}}  \right)^{3/2}\left(\frac{m}{\Gamma}  \right)^{1/4}\left(\frac{m}{M_{pl}}  \right)^2 M_{pl},
\end{eqnarray}
where $g_*$ is the number of degrees of freedom. Now, for the values of the masses involved in the process, the reheating temperature  is of the order

\begin{eqnarray}
T_{reh}\cong 3.5\times 10^{-18} h^{-1/2} g_*^{-1/4}M_{pl}\cong 8  h^{-1/2}  g_*^{-1/4} \mbox{ GeV},
\end{eqnarray}
which, for the number of the degrees of freedom for the energy density in the Standard Model, i.e. $g_*=106.75$, ranges between  $4$ GeV and $330$ TeV.

\

To end this subsection, we deal with the  second potential, i.e., with equation (\ref{PV2}),  which has  a smoother phase transition compared to  the first potential (\ref{PV}).
As we have already showed, in this case the energy density of the produced massive particles is given by 
\begin{eqnarray}
 \langle\rho(\tau)\rangle
 \cong 8\times 10^{-6}\left(\frac{m}{m_{\chi}}\right)^4 m^4\left(\frac{a_{kin}}{a(\tau)} \right)^3, 
\end{eqnarray}
and for the same decaying rate as in the previous cases 
the corresponding energy densities at the end of decay will be
\begin{eqnarray}
\rho_{\varphi, dec}=3{\Gamma}^2M_{pl}^2, \quad \mbox{ and } \quad \langle\rho_{dec}\rangle\cong 1.3\times10^{-5}\left(\frac{m}{m_{\chi}}  \right)^4{\Gamma}m^3.
\end{eqnarray}
Assuming, once again,  that the end of the decay occurs  before the radiation-domination epoch (i.e., $\langle \rho_{dec}\rangle\leq \rho_{\varphi, dec}$), one obtains the relation
\begin{eqnarray}\label{constraint0}
h^2\geq \frac{11\pi}{3}\times 10^{-5}\left(\frac{m}{m_{\chi}}  \right)^5\left(\frac{m}{M_{pl}}  \right)^2,
\end{eqnarray}
which for the values $m\cong 5\times 10^{-6} M_{pl}$ and {{} $m_{\chi}\cong 5\times 10^{-4} M_{pl}$ leads to the constraint $h\geq 5.3\times 10^{-13}$.}
On the other hand,  together with the condition $\Gamma\leq H_{kin}$, it leads to, $5.3\times 10^{-13}\leq h\leq 3.8\times 10^{-1}$.

Finally, if the thermalization of the relativistic plasma is instantaneous, the reheating temperature turns out to be 
\begin{eqnarray}
&&T_{reh}= \left(\frac{30}{\pi^2g_*} \right)^{1/4}\langle\rho_{dec}\rangle^{1/4}\sqrt{\frac{\langle\rho_{dec}\rangle}{\rho_{\varphi,dec}}}\cong 6.6\times 10^{-4} 
\left( \frac{m}{m_{\chi}} \right)^{\frac{13}{4}}\left( \frac{m}{M_{pl}} \right)^{2} g_*^{-1/4}h^{-1/2}M_{pl}\nonumber\\
&& \cong 5.2\times 10^{-21} g_*^{-1/4}h^{-1/2}M_{pl}\cong 12  g_*^{-1/4}h^{-1/2} \mbox { MeV},
\end{eqnarray}
which for  $g_*=106.75$  ranges between    $6$ MeV and $5$ TeV.

\subsection{Decay after the end of kination}

Now we assume that the decay of the $\chi$-field is after the end of kination.
Then, one  has to impose ${\Gamma}\leq H(\tau_r)\equiv H_r$, where we have denoted by $\tau_r$ the time at which kination ends. Taking this into account, one has 
\begin{eqnarray}\label{31}
H^2_{r}=\frac{2\rho_{\varphi, r}}{3M_{pl}^2}\quad \mbox{and} \quad \rho_{\varphi, r}=\rho_{\varphi, kin}\left( \frac{a_{kin}}{a_{r}} \right)^6=3H^2_{kin}M_{pl}^2\Theta^2,
\end{eqnarray}
in which, taking into account that during kination the energy density of the inflaton field decays as $a^{-6}$ and the one of the produced particles as $a^{-3}$,  we have introduced the so-called {\it heating efficiency}, defined as $\Theta\equiv \left( a_{kin}/a_{r} \right)^3=
\frac{\langle\rho_{kin}\rangle}{\rho_{\varphi, kin}}$. Consequently, from equation (\ref{31}), one can easily have $H_{r}=\sqrt{2}H_{kin}\Theta$ and, since
\begin{eqnarray}
\Theta=\left\{\begin{array}{cc}
2.3\times 10^{-20},& \mbox{for potential 1},\\
1.1\times 10^{-21},& \mbox{for potential 2},
\end{array}\right.\end{eqnarray}
one obtains that the parameter $h$ has to be
very small satisfying $h\leq 4.6\times 10^{-1}\sqrt{\Theta}$, which means that for the first potential $h\leq 7\times 10^{-11}$ while for the second potential, $h\leq 1.5\times 10^{-11}$.  Assuming once again the instantaneous thermalization, the reheating temperature (i.e., the temperature of the universe when the thermalized plasma starts to dominate) becomes
\begin{eqnarray}
T_{reh}=\left( \frac{30}{\pi^2 g_*} \right)^{1/4}\langle\rho_{dec}\rangle^{1/4}= \left( \frac{90}{\pi^2 g_*} \right)^{1/4}\sqrt{{\Gamma}M_{pl}}~,
\end{eqnarray}
where we have used that after  ${\tau}_r$ the energy density of the produced particles dominates the one  of the inflaton field. 
Then, we will have 

\begin{eqnarray}\label{decayafter}
T_{reh}\cong 7\times 10^{-3} hg_*^{-1/4} M_{pl}.
\end{eqnarray}
Consequently,  assuming that the BBN epoch occurs at the $1$ MeV regime 
and taking $g_*=106.75$, one can find that the value of $h$ resides in the interval $10^{-19}\lesssim h\lesssim10^{-11}$.

\section{Production of gravitational waves}
\label{sec-Gws}

In this Section we study the production of gravitational waves (GWs), which is the same as the gravitational particle production of massless particles minimally coupled to gravity, due to a sudden phase transition from
a de Sitter phase to an exact kination regime, i.e., when the EoS parameter is exactly $1$.

The model is given by the following dynamics. The conformal Hubble parameter for this model evolves as
 \begin{eqnarray}
 {\mathcal H}(\tau)=\left\{ \begin{array}{ccc}
 -\frac{1}{\tau}&   \mbox{for} &    \tau<\tau_{kin}<0\\
 &   &\\
 \frac{1}{(2\tau-3\tau_{kin})}&   \mbox{for} &            \tau\geq \tau_{kin},
 \end{array}\right.
 \end{eqnarray}
 and the scale factor evolves with 
 \begin{eqnarray}
 a(\tau)=\left\{ \begin{array}{ccc}
 -\frac{1}{H_{kin}\tau}&   \mbox{for} &            \tau<\tau_{kin}<0\\
 &   &\\
 a_{kin}\sqrt{\frac{2\tau-3\tau_{kin}}{-\tau_{kin}}}&   \mbox{for} &            \tau\geq \tau_{kin},
 \end{array}\right.
 \end{eqnarray}
 where $H_{kin}$ is the value of the Hubble parameter during the de Sitter phase and $a_{kin}=-\frac{1}{H_{kin}\tau_{kin}}.$ 
 The $k$-mode is given  by
 \begin{eqnarray}
 \chi_k(\tau)=\left\{ \begin{array}{ccc}
 \frac{1}{\sqrt{2k}} e^{-ik\tau}   \left(1-\frac{i}{k\tau}\right)&   \mbox{for} &            \tau<\tau_{kin}<0\\
 \alpha_k\sqrt{\frac{\pi(\tau-\frac{3}{2}\tau_{kin}  )}{4}}H^{(2)}_0\left(k(\tau-\frac{3}{2}\tau_{kin} ) \right)+
  \beta_k\sqrt{\frac{\pi(\tau-\frac{3}{2}\tau_{kin}  )}{4}}
  H^{(1)}_0\left(k(\tau-\frac{3}{2}\tau_{kin} )\right)
  &   \mbox{for} &            \tau\geq \tau_{kin},     
  \end{array}\right.
 \end{eqnarray}
 where $H_0^{(1)}$ and $H_0^{(2)}$  are the Hankel's functions. These modes satisfy the equation 
 \begin{eqnarray}
\chi''_k+\Omega_k^2(\tau) \chi_k=0,
\end{eqnarray}
where  we have introduced the notation $\Omega_k^2(\tau)\equiv k^2-\frac{a''}{a}$.

From a simple calculation one could find that $\frac{a''}{a}\propto a^2H^2$, which shows that the modes well inside the Hubble radius ($k\gg aH={\mathcal H}\propto \frac{1}{\tau}$) do not feel gravity and, thus, no particles are produced during the phase transition. So, only the ones well outside of the Hubble radius have to be used to compute the energy density of the produced particles, which is actually given by \cite{Bunch}
 \begin{eqnarray}
\langle\rho_{GW}(\tau)\rangle=\frac{1}{4\pi^2 a^4(\tau)}\int_0^{{\mathcal H}_{kin}}\left\{ (|\chi'_k|^2+k^2|\chi_k|^2-{k})-\left[ {\mathcal H}( |\chi_k|^2)' -{\mathcal H}^2|\chi_k|^2  \right]   \right\}k^2dk\nonumber\\
=\frac{1}{4\pi^2 a^2(\tau)}\int_0^{{\mathcal H}_{kin}}\left(\left|\left(\frac{\chi_k(\tau)}{a(\tau)}\right)'\right|^2+k^2\left|\frac{\chi_k(\tau)}{a(\tau)}\right|^2-{k}\right)k^2dk,
\end{eqnarray} 
where as in the massive case, the zero-point oscillations of the vacuum have been substracted. 

The calculation has to be done in three steps: 
\begin{enumerate}
\item For modes that are outside the Hubble radius at the beginning of kination and re-enter it during kination, i.e., satisfying  ${\mathcal H}_{r}<k< {\mathcal H}_{kin}$
(where we have denoted by ${\mathcal H}_r$ the value of the conformal Hubble parameter at the end of kination),  
when $\tau\gtrsim \tau_r$ one has $\frac{1}{2}\lesssim \frac{\tau}{2\tau_r}\cong \tau {\mathcal H}_r<k\tau$, so the modes 
practically do not feel gravity and, thus, we can make the approximation
\begin{eqnarray}
\chi_k(\tau)=\alpha_k
\frac{e^{-ik\tau}}{\sqrt{2k}}
+\beta_k\frac{e^{ik\tau}}{\sqrt{2k}}.
\end{eqnarray}
\item For modes that are outside of the Hubble radius at the end of kination ($k<{\mathcal H}_r$), we can use the small argument approximation of Hankel's functions
and  obtain
\begin{eqnarray}
\chi_k(\tau)=\alpha_k\sqrt{\frac{\pi(\tau-\frac{3}{2}\tau_{kin})}{4}}\left(
 1-\frac{2i}{\pi}\left(\gamma+\ln\left(\frac{k(\tau-\frac{3}{2}\tau_{kin})}{2}\right)\right) \right)\nonumber\\
+ \beta_k\sqrt{\frac{\pi(\tau-\frac{3}{2}\tau_{kin})}{4}}\left(
 1+\frac{2i}{\pi}\left(\gamma+\ln\left(\frac{k(\tau-\frac{3}{2}\tau_{kin})}{2}\right)\right) \right).
\end{eqnarray}
\item As we have already explained the relevant modes satisfy $k < {\mathcal H}_{kin} \Longleftrightarrow k|\tau_{kin}|<1$. Thus, in order to calculate the Bogoliubov coefficents, which are  obtained matching the modes at its first derivative at $\tau_{kin}$, one can use the small argument approximation of Hankel's functions and obtain that 
 \begin{eqnarray}
 \alpha_k=\frac{ie^{-ik\tau_{kin}}}{\sqrt{\pi}}\Bigg[\left(\frac{{\mathcal H}_{kin}}{k}\right)^{3/2}+\frac{1}{2}\left(\frac{{\mathcal H}_{kin}}{k}\right)^{-1/2}{ \left(\gamma+\ln\left(\frac{k}{4{\mathcal H}_{kin}} \right) \right)} \nonumber\\-i
 \left(  \left(\frac{{\mathcal H}_{kin}}{k}\right)^{1/2}+\frac{\pi}{4} \left(\frac{{\mathcal H}_{kin}}{k}\right)^{-1/2} \right)
   \Bigg]
 \end{eqnarray}
 
 \begin{eqnarray}
 \beta_k=\frac{ie^{-ik\tau_{kin}}}{\sqrt{\pi}}\Bigg[\left(\frac{{\mathcal H}_{kin}}{k}\right)^{3/2}+\frac{1}{2}\left(\frac{{\mathcal H}_{kin}}{k}\right)^{-1/2}{ \left(\gamma+\ln\left(\frac{k}{4{\mathcal H}_{kin}} \right) \right)}\nonumber\\-i
 \left(  \left(\frac{{\mathcal H}_{kin}}{k}\right)^{1/2}-\frac{\pi}{4} \left(\frac{{\mathcal H}_{kin}}{k}\right)^{-1/2} \right)
   \Bigg].
 \end{eqnarray} 
 
 Note that the Bogoliubov coefficients satisfy the well known relation $|\alpha_k|^2-|\beta_k|^2=1$ and the leading term of $\beta_k$ is $\frac{i}{\sqrt{\pi}}\left(\frac{{\mathcal H}_{kin}}{k}\right)^{3/2}$.

 \end{enumerate}

For modes  satisfying ${\mathcal H}_r<k<{\mathcal H}_{kin}$, the contribution to the energy density when $\tau\gtrsim \tau_r$ is
\begin{eqnarray}\label{A}
\frac{1}{2\pi^2 a^4(\tau)}\int_{{\mathcal H}_r}^{{\mathcal H}_{kin}}k^3|\beta_k|^2dk-
\frac{1}{4\pi^2 a^4(\tau)}\int_{{\mathcal H}_r}^{{\mathcal H}_{kin}}\left[ {\mathcal H}( |\chi_k|^2)' -{\mathcal H}^2|\chi_k|^2  \right]   k^2dk.
\end{eqnarray}

The first term leads to $\frac{1}{2\pi^3}H_{kin}^4\left(\frac{a_{kin}}{a(\tau)} \right)^4$ and the second one is bounded by
$\frac{1}{4\pi^2 a^4(\tau)}\int_{{\mathcal H}_r}^{{\mathcal H}_{kin}}[k {\mathcal H}+
{\mathcal H}^2]  (|\alpha_k|^2+|\beta_k|^2) kdk.$ Then, taking the leading terms of the Bogoliubov coefficients, one gets
\begin{eqnarray}
\frac{1}{4\pi^2 a^4(\tau)}\left|\int_{{\mathcal H}_r}^{{\mathcal H}_{kin}}\left[ {\mathcal H}( |\chi_k|^2)' -{\mathcal H}^2|\chi_k|^2  \right]   k^2dk\right|
\leq \frac{3}{2\pi^3}H_{kin}^4\ln\left(\frac{a_r}{a_{kin}}\right)
\left(\frac{a_{kin}}{a(\tau)} \right)^6+ \frac{1}{2\pi^3}H_{kin}^4\frac{a_{kin}^6a_r^2}{a^8(\tau)}
\end{eqnarray}
and, taking into account the bounds $\left(\frac{a_{kin}}{a_r} \right)\ll 1$ and $\ln\left(\frac{a_r}{a_{kin}}\right)\left(\frac{a_{kin}}{a_r} \right)^2\ll 1$, one can see that
the first term of (\ref{A}) is the leading one and its contribution to the energy density of GWs is $\frac{1}{2\pi^3}H_{kin}^4\left(\frac{a_{kin}}{a(\tau)} \right)^4$.

\

For modes satisfying $k<{\mathcal H}_r$, using the small argument approximation  and the formulas
\begin{eqnarray}
\frac{\alpha_k\chi_{k}(\tau)+
 \beta_k \chi_{k}^*(\tau)}{a(\tau)}=\frac{e^{-ik\tau_{kin}}}{a_{kin}\sqrt{2{\mathcal H}_{kin}}}
 \left[\left(\frac{{\mathcal H}_{kin}}{k} \right)^{1/2}+i\left\{\left(\frac{{\mathcal H}_{kin}}{k}\right)^{3/2} + 
 \frac{1}{2}\left(\frac{{\mathcal H}_{kin}}{k}\right)^{-1/2}\ln\left(\frac{{\mathcal H}}{{\mathcal H}_{kin}} \right) \right\} \right]
\end{eqnarray}
 and
 \begin{eqnarray}
\left( \frac{\alpha_k\chi_{k}(\tau)+
 \beta_k \chi_{k}^*(\tau)}{a(\tau)} \right)'=\frac{-ie^{-ik\tau_{kin}}}{2a_{kin}\sqrt{2{\mathcal H}_{kin}}}
 \frac{1}{\tau-\frac{3}{2}{\tau}_{kin}}\left(\frac{{\mathcal H}_{kin}}{k}\right)^{-1/2}=\frac{-ie^{-ik\tau_{kin}}}{2}\left(\frac{a_{kin}}{a}\right)^2\sqrt{\frac{2H_{kin}}{a_{kin}}}\left(\frac{{\mathcal H}_{kin}}{k}\right)^{-1/2},
 \end{eqnarray}
one obtains the following contribution to the energy density,
\begin{eqnarray}
\frac{H_{kin}^4}{32\pi^2}\left(\frac{a_{kin}^{14}}{a^6(\tau)a_r^8}+2\frac{a_{kin}^{6}}{a^2(\tau)a_r^4}
+\left(1+2\ln\left(\frac{a_{kin}}{a(\tau)} \right)\right)\frac{a_{kin}^{10}}{a^2(\tau)a_r^8}
+\frac{1}{3}{H_{kin}^4}\ln^2\left(\frac{a_{kin}}{a(\tau)} \right)\frac{a_{kin}^{14}}{a^2(\tau)a^{12}_{r}}\right),\end{eqnarray}
which is  sub-leading compared to $\frac{1}{2\pi^3}H_{kin}^4\left(\frac{a_{kin}}{a(\tau)} \right)^4$ and, thus, one can conclude that
the energy density of  GWs when $\tau\gtrsim \tau_r$ turns out to be 
\begin{eqnarray}
\langle\rho_{GW}(\tau)\rangle\cong \frac{H_{kin}^4}{2\pi^3}\left( \frac{a_{kin}}{a(\tau)}\right)^4\cong 
10^{-2} H_{kin}^4\left( \frac{a_{kin}}{a(\tau)}\right)^4.
\end{eqnarray}

{ We close this section with a short remark on the $\beta$-Bogoliubov coefficient. We noted that the $\beta$-Bogoliubov coefficient calculated by us mildly differs 
from \cite{Giovannini99}. In particular, eqn. (C.2) of Appendix C of \cite{Giovannini99} has a very mild mismatch with us. However, such difference does not affect the main results and conclusions of \cite{Giovannini99} apart from a factor in the BBN bound. However, inspite of that for interested readers we present our calculations in Appendix \ref{appx-C}. }

\subsection{When does the overproduction of GWs not affect the  BBN success?}
\label{sec-overproduction}

The success of the BBN demands that the ratio of the energy density of GWs to the one of the produced particles at the reheating time satisfies
\cite{dimopoulos}
\begin{eqnarray}\label{bbnconstraint}
\frac{\langle\rho_{GW, reh}\rangle}{\langle\rho_{reh}\rangle}\leq 10^{-2}.
\end{eqnarray} 

This bound could never be accomplished when reheating 
is due to the gravitational production of massless particles because the energy density of those particles decreases as 
the one of GWs \cite{pv}, i.e., 
as we have already seen in the previous section, close to the end of kination the energy density decreases as $ 10^{-2} H^4_{kin} \left(a_{kin}/a(\tau) \right)^4$ .

In the same way, dealing with heavy massive particles,
first of all we see that the constraint (\ref{bbnconstraint}) is never overpassed when the decay of the massive particles is previous to the end of  kination. Effectively,
if the decay occurs after  the end of kination one can calculate 
$\frac{\langle\rho_{GW}(\tau)\rangle}{\langle\rho(\tau)\rangle}$ at the end of kination.
Precisely, using equation (\ref{31}) and the fact that $\Theta=\left(a_{kin}/a_{r}  \right)^3$, one finds
\begin{eqnarray}
\frac{\langle\rho_{GW, r}\rangle}{\langle\rho_{r}\rangle}= \frac{1}{3}10^{-2} \left(\frac{H_{kin}}{M_{pl}}\right)^2\Theta^{-2/3}\cong 
\left\{\begin{array}{cc}
3.7\times 10^{-1}~,& \mbox{for potential 1}, \\
2.8~, & \mbox{ for potential 2}
.\end{array}\right.
\end{eqnarray}

This result shows that, if the decay occurs before the end of kination, the constraint (\ref{bbnconstraint}) is never achieved because after the decay the energy density of the produced particles decreases as the one of the GWs, so 
in that case 
$\frac{\langle\rho_{GW, reh}\rangle}{\langle\rho_{reh}\rangle}$ is greater than   $3.7\times 10^{-1} $ for the first potential and it is also greated than $2.8$ for the second one.

Hence, in order to overpass the constraint, the decay must be produced after the end of kination. And, assuming once again the instantaneous thermalization,  the reheating time will coincide with the decay one. Then, since
$\langle\rho_{ dec}\rangle=3{\Gamma}^2M_{pl}^2$ and 
\begin{eqnarray}
H_{dec}=H_{r}\left( \frac{a_{r}}{a_{dec}} \right)^{3/2}\Longrightarrow \left( \frac{a_{r}}{a_{dec}} \right)^{3/2}=\frac{\Gamma}{\sqrt{2}H_{kin}\Theta},
\end{eqnarray}
we  will have 
\begin{align}
\langle\rho_{GW, dec}\rangle=\langle\rho_{GW, r}\rangle\left( \frac{a_{r}}{a_{dec}} \right)^4= 
\langle\rho_{GW, r}\rangle\left( \frac{\Gamma}{\sqrt{2}H_{kin}\Theta}    \right)^{8/3}
=10^{-2} H^4_{kin}\Theta^{-4/3}\left( \frac{\Gamma}{\sqrt{2}H_{kin} }  \right)^{8/3},
\end{align}
and thus,
\begin{eqnarray}
\frac{\langle\rho_{GW, reh}\rangle}{\langle\rho_{reh}\rangle}\cong 10^{-4} \left(\frac{h}{\Theta}  \right)^{4/3}\frac{m_{\chi}^{2/3}H^{4/3}_{kin}}{M_{pl}^2}\cong 
\left\{\begin{array}{cc}
4\times 10^{12} h^{4/3}~,&\mbox{ for potential 1}\\
2.4\times 10^{14} h^{4/3}~,&\mbox{ for potential 2}, 
\end{array}\right.\end{eqnarray}
from which one can see that the constraint (\ref{bbnconstraint}) is satisfied for $h\leq 1.1\times 10^{-11}$ (for the first potential) and for  $h\leq 5\times 10^{-13}$ (for the second potential). 
Therefore,  for $g_*=106.75$ and using the equation (\ref{decayafter}), one can see that the maximum reheating temperature in the case of the first potential turns out to b, $T_{reh}\cong 57$ TeV, while for the second potential $T_{reh}\cong 3$ TeV.

\

{ A final remark is in order:
After  the discovery of the Higgs  boson, it is well-know that there exists at least one other scalar field, which during inflation it appears to be a spectator field with no dynamical role \cite{Enqvist}. The Standard Model Higgs doublet could be parametrized with a single scalar degree of freedom, namely $\phi$, whose potential for large amplitudes is just given by a quadratic potential \cite{Figueroa}
\begin{eqnarray}
V(\phi)=\frac{\lambda}{4}\phi^4,
\end{eqnarray}
where $\lambda$ is the self-coupling constant.

\

It has been showed in section 2.1 of  \cite{Enqvist} (see also section II A of \cite{Figueroa}) that at the end of inflation the energy density of the Higgs field, namely
$\rho_{\phi}$,  is $\rho_{\phi}\sim 10^{-3} H_*^4$, where  $H_*$ is the Hubble scale at the end of inflation \cite{Freese}, that is, $H_*=H_{end}\cong H_{kin}$, because 
there is not substantial drop of energy between the end of inflation and the beginning of kination. So, at the beginning of kination  the energy density of the Higgs scalar is approximately one order less than the energy density of the GWs (see formula (40)).

\

On the other hand, assuming that the Higgs field starts to oscillate immediately after the end of inflation,  then since the potential is quartic, using the Virial Theorem we can deduce that during the oscillations its effective Equation of State parameter is given by $w_{eff}=1/3$ \cite{Turner}, thus,  its energy density decays as radiation.
Therefore, since its decay product are light particles one can conclude that $\rho_{\phi}(\tau)\leq \langle\rho_{GW}(\tau)\rangle$ after the beginning of kination. This means that at the reheating time 
\begin{eqnarray}
\frac{\rho_{\phi, reh}}{\langle\rho_{reh}\rangle}\leq
\left\{\begin{array}{cc}
4\times 10^{12} h^{4/3}~,&\mbox{ for potential 1}\\
2.4\times 10^{14} h^{4/3}~,&\mbox{ for potential 2}, 
\end{array}\right.\end{eqnarray}
and for a very low reheating temperature, for example $1$ MeV, which corresponds to $h\sim 10^{-19}$ (see below formula (25)), one gets
\begin{eqnarray}
\frac{\rho_{\phi, reh}}{\langle\rho_{reh}\rangle}\leq
\left\{\begin{array}{cc}
1.8\times 10^{-13} ~,&\mbox{ for potential 1}\\
1.1\times 10^{-11} ~,&\mbox{ for potential 2},
\end{array}\right.\end{eqnarray}
that is, at the reheating time the energy density from the Higgs condensate decay is completely negligible compared with the energy density of the decay products
of the superheavy $\chi$-field.

}

\section{Conclusions}

The description of both early inflationary phase and late quintessence phase in a single  framework was named as quintessential inflationary models by Peebles and Vilenkin. 
This class of unified cosmic models has gained a robust attention to the cosmological community since its appearance. Later on, the developments of the observational data have clarified many issues, including the shortcomings of those models, and eventually the quintessential inflationary models have been revised either by replacing the inflationary piece of the models or by introducing a different reheating mechanism via gravitatational particle production. The present work has aimed to discuss the understanding of the gravitational particle production in such models.

Thus, assuming two quintessential inflationary models, we study the creation of superheavy massive particles conformally coupled to gravity at the beginning of kination regime, where the adiabatic regime is broken. First of all we have shown how to perform the calculation of the energy density of the produced particles using the well-known diagonalization method, proving that before the beginning of kination the one-loop energy density of the vacuum only contains sub-dominant geometric polarization terms, i.e., terms that do not affect the classical Friedmann equation. Only after the beginning of kination, where the adiabatic regime is broken, particles are created and its energy density is calculated.

 We also show that the same energy density of the produced particles could be obtained by approximating the vacuum modes using the WKB approximation and performing the
 matching of the modes at its first  derivative
at the beginning of kination.  Since these superheavy  particles have to decay in lighter ones to form a relativistic plasma which eventually becomes dominant and matches with the hot big bang universe, two different situations arise, namely,  when the decay occurs  before the end of kination regime and when the decay occurs after the end of the kination regime. Thus,  for both situations we have calculated the reheating temperature of the universe, i.e., the temperature of the universe when the energy density of the inflaton field is of the same order as  the relativistic plasma as a function of the decay rate.

Finally, we have also reviewed with all the details the calculation of  the energy density of the produced GWs due to the phase transition from inflation to kination, 
obtaining a $\beta$-Bogoliubov coefficient differing by a logarithmic term \cite{Giovannini99}. Such a difference plays no effective role because, apart from a numerical factor in the BBN bound, nothing actually changes. 
Moreover, we have also shown that, in order that this overproduction of GWs does not affect  the BBN success, the decay of the heavy massive particles must be after the end of kination, obtaining reheating temperatures in the TeV regime.

\section*{Acknowledgments}

This investigation has been supported by MINECO (Spain) grants  MTM2014-52402-C3-1-P and MTM2017-84214-C2-1-P, and  in part by the Catalan Government 2017-SGR-247. SP acknowledges the research grant under Faculty Research and Professional Development Fund (FRPDF) Scheme of Presidency University, Kolkata, India.  The authors thank Prof. M. Giovannini and Prof. J. D. Barrow for useful correspondence.

\appendix

\section{The diagonalization method}
\label{sec-diagonalization}

The diagonalization method was developed during the seventies of last century by the Russian scientists Grib, Frolov, Mamayev, Mostepanenko \cite{fmm,glm,gmm}
and also by Zeldovich and Starobinsky \cite{zs}. Principally, for a  quantum scalar field  of superheavy  particles conformally coupled to gravity, 
namely $\chi$, the Klein-Gordon (K-G) equation in the flat Friedmann-Lema{\^\i}tre-Robertson-Walker (FLRW) spacetime follows 
\begin{eqnarray}\label{kg0}
\chi''+2{\mathcal H}\chi'- \nabla^2\chi+\left(m_{\chi}^2a^2+\frac{a''}{a}    \right)\chi=0,
\end{eqnarray}
where the prime attached to any quantity denotes the derivative with respect the conformal time $\tau$; ${\mathcal H} \equiv a'/a$, is the conformal Hubble parameter and $m_{\chi}$ is the mass of the scalar field. Now, writing the quantum field in Fourier space,
\begin{eqnarray}
\chi({\bf x},\tau)=\frac{1}{(2\pi)^{3/2}a}\int d^3k\left( \hat{a}_{\bf{k}}\chi_{ k}(\tau) e^{-i\bf{k}.\bf{x}}+ 
 \hat{a}_{\bf{k}}^{\dagger}\chi_{ k}^*(\tau) e^{i\bf{k}.\bf{x}} \right),
\end{eqnarray}
 where $d^3 k=dk_1dk_2dk_3$, ${\bf k}=(k_1,k_2,k_3)$, ${\bf x}=(x_1,x_2,x_3)$,   $k=\sqrt{k_1^2+k^2_2+k_3^2}$ and $\hat{a}_{\bf k}$ is the annihilation operator corresponding to the vacuum state at a given initial time $\tau_i$,
 which is defined by the condition
 \begin{eqnarray}
 \chi_{k}(\tau_i)=
 \frac{1}{\sqrt{2\omega_k(\tau_i)}}e^{-i\int^{\tau_i} \omega_k(\bar\eta)d\bar\eta}, \quad
 \chi_{ k}'(\tau_i)=
-i \omega_k(\tau_i)\chi_{ k}(\tau_i), \end{eqnarray}
 with $\omega_k(\tau)=\sqrt{k^2+m_{\chi}^2a^2(\tau)}$,
the Klein-Gordon equation (\ref{kg0}) becomes
\begin{eqnarray}\label{kg1}
\chi_{ k}''(\tau)+\omega^2_k(\tau) \chi_{ k}(\tau)=0,
\end{eqnarray}
which is the equation of a harmonic oscillator with 
time dependent frequency $\omega_k(\tau)$. Additionally, the energy density of the vacuum is given by \cite{Bunch}
\begin{eqnarray}\label{vacuum-energy}
\langle\rho(\tau)\rangle\equiv \langle 0| \hat{\rho}(\tau)|0 \rangle=
\frac{1}{4\pi^2a^4(\tau)}\int_0^{\infty} k^2dk \left(   |\chi_{ k}'(\tau)|^2+ \omega^2_k(\tau) |\chi_{ k}(\tau)|^2-  \omega_k(\tau)        \right),
\end{eqnarray}
where in order to obtain a finite energy density \cite{gmmbook} we have subtracted the energy density of the zero-point oscillations of the vacuum 
$\frac{1}{(2\pi)^3a^4(\tau)}\int d^3k  \frac{1}{2} \omega_k(\tau)$.

\begin{remark}
 For a quantum field  not conformally coupled to gravity, it is not enough to subtract the energy density of the zero-point oscillations of the vacuum to get
 a finite energy density. In that case one needs a more complicated regularization process such as the subtraction of  adiabatic terms up to the four order \cite{Bunch},
  the point splitting method \cite{bunch, Birrell} or the  $n-$wave procedure \cite{gmmbook}.
\end{remark}

We follow the method developed in \cite{zs} (see also Section $9.2$ of \cite{gmmbook}), hence we write 
\begin{eqnarray}\label{zs}
\chi_{k}(\tau)= \alpha_k(\tau)\frac{e^{-i\int^{\tau} \omega_k(\bar\eta)d\bar\eta}}{\sqrt{2\omega_k(\tau)}}+
\beta_k(\tau)\frac{e^{i\int^{\tau} \omega_k(\bar\eta)d\bar\eta}}{\sqrt{2\omega_k(\tau)}},\end{eqnarray}
where $\alpha_k(\tau)$ and $\beta_k(\tau)$ are the time-dependent Bogoliubov coefficients.
Now, imposing that the modes satisfy   the condition
\begin{eqnarray}
\chi_{k}'(\tau)= -i\omega_k(\tau)\left(\alpha_k(\tau)\frac{e^{-i\int^{\tau} \omega_k(\bar\eta)d\bar\eta}}{\sqrt{2\omega_k(\tau)}}-
\beta_k(\tau)\frac{e^{i\int^{\tau} \omega_k(\bar\eta)d\bar\eta}}{\sqrt{2\omega_k(\tau)}}\right),\end{eqnarray}
one can show that   the Bogoliubov coefficients must satisfy the system 
\begin{eqnarray}\label{Bogoliubovequation}
\left\{ \begin{array}{ccc}
\alpha_k'(\tau) &=& \frac{\omega_k'(\tau)}{2\omega_k(\tau)}e^{2i\int^{\tau} \omega_k(\bar\eta)d\bar\eta}\beta_k(\tau)\\
\beta_k'(\tau) &=& \frac{\omega_k'(\tau)}{2\omega_k(\tau)}e^{-2i\int^{\tau}\omega_k(\bar\eta)d\bar\eta}\alpha_k(\tau),\end{array}\right.
\end{eqnarray}
and thus the expression (\ref{zs}) is the solution of the equation (\ref{kg1}).

\begin{remark}
Since the Wronskian is conserved and $W[\chi_k(\tau_i),\chi^*_k(\tau_i)]\equiv \chi_k(\tau_i)(\chi^*_k)'(\tau_i)
-\chi_k'(\tau_i)\chi^*_k(\tau_i)
=i$, one can see that the Bogoliubov coefficients satisfy the equation $|\alpha_k(\tau)|^2- |\beta_k(\tau)|^2=1$.
\end{remark}

Finally, inserting (\ref{zs}) into the expression for vacuum energy density (\ref{vacuum-energy}), one finds that
\begin{eqnarray}
\langle\rho(\tau)\rangle= \frac{1}{2\pi^2a^4(\tau)}\int_0^{\infty} k^2\omega_k(\tau)|\beta_k(\tau)|^2 dk.
\end{eqnarray}

Coming back to the equation  (\ref{Bogoliubovequation}), 
in the first approximation taking $\alpha_k(\tau)=1$, 
we get
\begin{eqnarray}
\beta_k(\tau)=\int^{\tau}\frac{\omega_k'(\eta)}{2\omega_k(\eta)}e^{-2i\int^{\eta} \omega_k(\bar\eta)d\bar\eta}d\eta.
\end{eqnarray}

Finally, it is important to stress   that the classical Friedmann equation is modified by the following semi-classical equation $H^2=\frac{1}{3M_{pl}^2}\left({\rho+\langle\rho\rangle}
\right)$.

\section{The use of the  WKB approximation to calculate particle production}
\label{sec-wkb}

The  Wentzel-Kramers-Brilloui (WKB) approximation  applied to cosmology (see for instance \cite{Winitzki, Haro}, and references therein) shows that  the vacuum mode during the adiabatic regime can be approximated by
\begin{eqnarray}
{\chi}_{n,k}^{WKB}(\tau)\equiv
\sqrt{\frac{1}{2W_{n,k}(\tau)}}e^{-{i}\int^{\tau}W_{n,k}(\eta)d\eta},
\end{eqnarray}
where $n$ is the order of the approximation and $W_{n,k}(\tau)$ is calculated as follows (see for more details \cite{Winitzki}). 
First of all, instead of equation (\ref{kg1}) we consider the following equation 
\begin{eqnarray}\label{kgepsilon}
\epsilon\bar{\chi}_k''+\omega_k^2(\tau){\chi}_k=0,
\end{eqnarray}
where $\bar\epsilon$ is a dimensionless parameter that one may set  $\bar\epsilon=1$ at the end of calculations. 
Looking for a solution of (\ref{kgepsilon}) of the form 
\begin{eqnarray}\label{wkb-choice}
{\chi}_{n,k}^{WKB}(\tau; \bar\epsilon)=
\frac{1}{\sqrt{2W_{n,k}(\tau;\bar\epsilon)}}e^{-\frac{i}{\bar\epsilon}\int^{\tau}W_{n,k}(\eta;\bar\epsilon)d\eta},
\end{eqnarray}
where $W_{0,k}(\tau;\bar\epsilon)\equiv \omega_k(\tau)$, 
inserting (\ref{wkb-choice}) into (\ref{kgepsilon}) and  collecting the terms of order $\bar\epsilon^{2n}$,  one arrives at the iterative formula
\begin{eqnarray}
W_{n,k}(\tau;\bar\epsilon)= \mbox{ terms up to order } \bar\epsilon^{2n} \mbox{ of } 
\left( \sqrt{\omega_k^2(\tau)-\bar\epsilon^2 \left[\frac{1}{2} \frac{W''_{n-1,k}(\tau;\bar\epsilon)}{W_{n-1,k}(\tau;\bar\epsilon)} 
-\frac{3}{4}\frac{(W'_{n-1,k}(\tau;\bar\epsilon))^2}{W^2_{n-1,k}(\tau;\bar\epsilon)} \right]}         \right).
\end{eqnarray}

For the first potential (\ref{PV}) one only needs the first order WKB solution to approximate  the $k$-vacuum modes before and after the beginning of kination, given by 
\begin{eqnarray}
{\chi}_{1,k}^{WKB}(\tau)\equiv
\sqrt{\frac{1}{2W_{1,k}(\tau)}}e^{-{i}\int^{\tau}W_{1,k}(\eta)d\eta},
\end{eqnarray}
where $W_{1,k}$ has the expression \cite{Winitzki}
\begin{eqnarray}
W_{1,k}=
\omega_k-\frac{1}{4}\frac{\omega''_{k}}{\omega^2_{k}}+\frac{3}{8}\frac{(\omega'_{k})^2}{\omega^3_{k}} ,
\end{eqnarray}
because $W_{1,k}$ contains the first derivative of the Hubble parameter and, since the matching involves the derivative of the mode
and the second derivative of the Hubble parameter is discontinuous at $\tau_{kin}$, 
 the $\beta$-Bogoliubov coefficient does not
vanish. Effectively, 
before the beginning of kination the  vacuum mode is depicted by $\chi_{1,k}^{WKB}(\tau)$, but after $\tau_{kin}$
 this mode becomes a mix of positive and negative frequencies of the form
$\alpha_k \chi_{1,k}^{WKB}(\tau)+\beta_k (\chi_{1,k}^{WKB})^*(\tau)$, which is the manifestation of the particle production.
The $\beta_k$-Bogoliubov coefficient is obtained matching both expressions at $\tau_{kin}$, leading to
\begin{eqnarray}
\beta_k(\tau)=\frac{{\mathcal W}[\chi_{1,k}^{WKB}(\tau_{kin}^-),\chi_{1,k}^{WKB}(\tau_{kin}^+)]}
{{\mathcal W}[(\chi_{1,k}^{WKB})^*(\tau_{kin}^+),\chi_{1,k}^{WKB}(\tau_{kin}^+)]}
= i{\mathcal W}[\chi_{1,k}^{WKB}(\tau_{kin}^-),\chi_{1,k}^{WKB}(\tau_{kin}^+)],
\end{eqnarray}
where ${\mathcal W}[f,g]=fg'-f'g$ denotes the Wronskian of the functions $f$ and $g$, 
and we have introduced the notation $f(\tau^+_{kin})= \lim_{\tau\rightarrow \tau_{kin}; \tau>\tau_{kin}}f(\tau)$ and
  $f(\tau^-_{kin})= \lim_{\tau\rightarrow \tau_{kin}; \tau<\tau_{kin}}f(\tau)$.
  
  \

The square modulus of the $\beta$-Bogoliubov coefficient will be given approximately by   \cite{hap}
\begin{eqnarray}
 |\beta_k(\tau)|^2\cong \frac{m^4_{\chi}a^{10}_{kin}\left(\ddot{H}(\tau_{kin}^+)-\ddot{H}(\tau_{kin}^-)\right)^2}{256\omega_k^{10}(\tau_{kin})},
\end{eqnarray}
which coincides with the square modulus of the  leading term of the integration constant $C$ obtained in equation (\ref{constant}), 
as happens with the second potential.
This shows the equivalence between the 
methods to obtain the  energy density of the produced particles.

\section{An additional remark on the $\beta$-Bogoliubov coefficient}
\label{appx-C}

In  Ref. \cite{Giovannini99}, the author obtains that the leading  value of the $\beta$-Bogoliubov coefficient is
$\beta_k\sim \frac{9}{4\sqrt{\pi}}\left(\frac{{\mathcal H}_{kin}}{k}\right)^{3/2}\ln\left(\frac{k}{{\mathcal H}_{kin}} \right)$. However, it seems to us that there might be a very mild change in the $\beta$-Bogoliubov coefficient which of course does not affect the main results and the conclusion of the paper apart from a factor in the BBN bound. Hence, there is absolutely no worry at all. 
We find that  the term containing $\left(\frac{{\mathcal H}_{kin}}{k}\right)^{3/2}\ln\left(\frac{k}{{\mathcal H}_{kin}} \right)$ vanishes and the leading term becomes
 $\frac{i}{\sqrt{\pi}}\left(\frac{{\mathcal H}_{kin}}{k}\right)^{3/2}$. Effectively, using the long wave-length approximation one has
 \begin{eqnarray}
&&\chi_k(\tau_{kin}^-)=\frac{-i}{\sqrt{2k}k\tau_{kin}};\quad  \chi_k'(\tau_{kin}^-)=i\sqrt{\frac{k}{2}}\frac{1}{k^2\tau^2_{kin}};\quad 
 \chi_k(\tau_{kin}^+)=-i\sqrt{\frac{-\tau_{kin}}{2\pi}} \ln\left(-\frac{k\tau_{kin}}{4}  \right);\nonumber \\
&&\chi_k'(\tau_{kin}^+)=-\frac{i}{\sqrt{-2\pi\tau_{kin}}} \ln\left(-\frac{k\tau_{kin}}{4}  \right)-i\sqrt{\frac{2}{-\pi\tau_{kin}}}.
  \end{eqnarray} 
    
Then, since $\beta_k=i{\mathcal W[}\chi_k(\tau^-);\chi_k(\tau^+)]$,  a simple calculation  proves our statement, i.e.,
  $\beta_k\cong \frac{i}{\sqrt{\pi}}\frac{1}{(-k\tau_{kin})^{3/2}}=\frac{i}{\sqrt{\pi}}\left(\frac{{\mathcal H}_{kin}}{k}\right)^{3/2}$. If one recalculates the computations done in \cite{Giovannini99} in order to obtain the $\beta$-Bogoliubov coefficient ($A_-(k)$ in its notation) one obtains the following expression:

\begin{align}
\beta_k=-\frac{\pi}{4\sqrt{2}}e^{-\frac{i}{2}\pi(\nu+1)}\left\{ H^{(2)}_0\left(\frac{x_1}{2} \right)\left[\frac{3}{2}H^{(2)}_{\nu}(-x_1)+\frac{x_1}{2}\left( H^{(2)}_{\nu+1}(-x_1)-H^{(2)}_{\nu-1}(-x_1)\right) \right] -x_1H^{(2)}_1\left(\frac{x_1}{2}\right)H^{(2)}_{\nu}(-x_1) \right\},
\end{align}
where $x_1=-k\tau_{kin}$ and $\nu=3/2$. We note that, when $|x_1|\ll 1$, $H^{(2)}_ {\nu-1}(-x_1)$ is subdominant relative to $H^{(2)}_{\nu+1}(-x_1)$. So, by ignoring this term, our expression almost coincides with the one in equation (C.2) in \cite{Giovannini99} with only difference of a minus sign in front of $H^{(2)}_{\nu}(-x_1)$. This minus sign appears to be important as we show next. By using the recurrence relation $\frac{2\alpha}{x}Z_{\alpha}(x)=Z_{\alpha-1}(x)+Z_{\alpha+1}(x)$, being $Z_{\alpha}$ any combination of Bessel functions of order $\alpha$, we find that $H^{(2)}_{\nu}(-x_1)=\frac{-x_1}{2\nu}\left(H^{(2)}_{\nu-1}(-x_1)+H^{(2)}_{\nu+1}(-x_1) \right)$. Therefore the dominant terms multiplying $H^{(2)}_0\left(\frac{x_1}{2} \right)$ get cancelled each other and, hence, the only remaining dominant term turns out to be


\begin{eqnarray}
\beta_k \sim \frac{\pi}{4\sqrt{2}}e^{-\frac{i}{2}\pi (\nu+1)}x_1H^{(2)}_1\left(\frac{x_1}{2} \right) H^{(2)}_{\nu}(-x_1)
\sim \frac{i}{\sqrt{\pi}}e^{-\frac{3i\pi}{4}}\frac{1}{(k\tau_{kin})^{3/2}}=
\frac{i}{\sqrt{\pi}}e^{-\frac{i\pi}{4}}\left(\frac{{\mathcal H}_{kin}}{k} \right)^{3/2},
\end{eqnarray}
and thus
$|\beta_k|^2\sim \frac{1}{\pi}\left(\frac{{\mathcal H}_{kin}}{k} \right)^3$.

From our viewpoint this mild mismatch in \cite{Giovannini99} may come from the fact that during the de Sitter phase the conformal time is negative. However, the author uses
the vacuum mode 
$e^{-i\pi\nu/2}e^{-i\pi/4} \sqrt{\frac{\pi\tau}{4}}H_{\nu}^{(2)}(k\tau)$ (see formula $(3.4)$ of \cite{Giovannini99}), which contains square roots of negative numbers that complicate the calculations instead of 
$e^{i\pi\nu/2}e^{i\pi/4} \sqrt{\frac{-\pi\tau}{4}}H_{\nu}^{(1)}(-k\tau)$,
which has a positive argument that facilitates the calculations, obtaining
\begin{eqnarray}
\beta_k={ i} \frac{\pi}{4\sqrt{2}}e^{\frac{i}{2}\pi\nu}\left\{ H^{(2)}_0\left(\frac{x_1}{2} \right)\left[\frac{3}{2}H^{(1)}_{\nu}(x_1)-\frac{x_1}{2}\left( H^{(1)}_{\nu+1}(x_1)-H^{(1)}_{\nu-1}(x_1)\right) \right] -x_1H^{(2)}_1\left(\frac{x_1}{2}\right)H^{(1)}_{\nu}(x_1) \right\} \nonumber\\
\sim -{ i} \frac{\pi}{4\sqrt{2}}e^{{ \frac{i}{2}\pi \nu}}x_1H^{(2)}_1\left(\frac{x_1}{2} \right) H^{(1)}_{\nu}(x_1)\sim 
\frac{1}{\sqrt{\pi}}e^{\frac{i\pi}{4}}\left(\frac{{\mathcal H}_{kin}}{k} \right)^{3/2}.
\end{eqnarray}

A consequence of such mild mismatch is that the energy density per logarithmic interval of longitudinal momentum, for ${\mathcal H}_r<k<{\mathcal H}_{kin}$, is now given by
  \begin{eqnarray} 
  \rho(k,\tau)=\frac{d\rho_{GW}(k,\tau)}{d\ln k}
  =\frac{kd\rho_{GW}(k,\tau)}{dk}= 
  \frac{k^4}{2\pi^2a^4(\tau)}|\beta_k|^2\cong
  \frac{1}{2\pi^3}H_{kin}^4
  \left(\frac{k}{{\mathcal H}_{kin}} \right)\left( \frac{a_{kin}}{a(\tau)}\right)^4,
   \end{eqnarray}
  which differs from a logarithmic term of the result obtained in formula $(3.31)$ of \cite{Giovannini}. Fortunately, this only affects by a factor of one half  the BBN bound \cite{Giovannini99}:
$\frac{h_0^2}{\rho_c(\tau_0)} \int_{{\mathcal H}_{BBN}}^{{\mathcal H}_{end}}
\rho(k,\tau_0)d\ln k \leq 10^{-5}$,
where $h_0$ parametrizes the experimental uncertainty to determine the current value
of the Hubble constant, $\rho_c(\tau_0)$ is the current value of the critical density and   ${\mathcal H}_{BBN}$ and   ${\mathcal H}_{end}$ are respectively the values of the
conformal Hubble parameter at the BBN and at the end of inflation, because, although one uses the  formula $(3.31)$ of \cite{Giovannini}, the logarithmic terms 
are all sub-dominant (see for instance \cite{rubio}).



\begin{thebibliography}{99}


\bibitem {guth}
A. Guth,
Phys. Rev. {\bf D 23}, 347 (1981).



\bibitem{linde}
A. Linde, 
Phys. Lett. {\bf B 108}, 389 (1982).


\bibitem{Burd:1988ss} 
  A.~B.~Burd and J.~D.~Barrow,
  Nucl.\ Phys.\ B {\bf 308}, 929 (1988).



\bibitem{Barrow:1990vx} 
  J.~D.~Barrow,
  Phys.\ Lett.\ B {\bf 235}, 40 (1990).

\bibitem{Barrow:1994nt} 
  J.~D.~Barrow,
  Phys.\ Rev.\ D {\bf 49}, 3055 (1994).


\bibitem{chibisov}
G. Chibisov and V. Mukhanov,   
Mon. Not. Roy. Astron. Soc.  {\bf 200}, 535 (1982).

\bibitem{starobinsky}
A. A. Starobinsky,
Phys. Lett. {\bf B 117}, 175 (1982).


\bibitem{pi}
A. H. Guth and S-Y. Pi, 
Phys. Rev. Lett. {\bf 49}, 1110  (1982).

\bibitem{bardeen}
J. M. Bardeen, P. J. Steinhardt and M. S. Turner,
Phys. Rev.  {\bf D 28}, 679  (1983).


\bibitem{Linde:1982uu} 
  A.~D.~Linde,
  Phys.\ Lett.\  {\bf 116 B}, 335 (1982).




\bibitem{Planck}
P.~A.~R.~Ade {\it et al.}  [Planck Collaboration],  
Astron.\ Astrophys.\  {\bf 594}, A20 (2016)
[arXiv:1502.02114].


\bibitem{Copeland:2006wr} 
  E.~J.~Copeland, M.~Sami and S.~Tsujikawa,
  Int.\ J.\ Mod.\ Phys.\ D {\bf 15}, 1753 (2006)
  [hep-th/0603057].

\bibitem{pv}
P. J. E. Peebles and  A. Vilenkin,
Phys. Rev. {\bf D 59}, 063505 (1999)  [arXiv:9810509].

{
\bibitem{dimopoulos01}
K. Dimopoulos,
Nucl. Phys. Proc. Suppl {\bf 95}, 70 (2001) [arXiv:astro-ph/0012298]

}


\bibitem{dimopoulos1}
K. Dimopoulos and J. W. F. Valle,
Astropart. Phys. {\bf 18}, 287 (2002) 	[arXiv:0111417].


\bibitem{Giovannini:2003jw} 
  M.~Giovannini,
  Phys.\ Rev.\ D {\bf 67}, 123512 (2003)
  [hep-ph/0301264].
  
  \bibitem{hossain1}
Md.~Wali Hossain, R. Myrzakulov, M. Sami and E. N. Saridakis, 
Phys. Rev. D {\bf 89}, 123513 (2014)  	[arXiv:1404.1445 [gr-qc]]
	

\bibitem{hossain3}
Md.~Wali Hossain, R.~Myrzakulov, M.~Sami and E.~N.~Saridakis,
  Int.\ J.\ Mod.\ Phys.\ D {\bf 24}, no. 05, 1530014 (2015)
  [arXiv:1410.6100 [gr-qc]].
  
 \bibitem{deHaro:2016hpl} 
  J.~de Haro, J. Amor\'{o}s and S.~Pan,
  Phys.\ Rev.\ D {\bf 93}, no. 8, 084018 (2016)
  [arXiv:1601.08175 [gr-qc]]. 
  
 \bibitem{deHaro:2016hsh} 
  J.~de Haro and E.~Elizalde,
  Gen.\ Rel.\ Grav.\  {\bf 48}, no. 6, 77 (2016)
  [arXiv:1602.03433 [gr-qc]].
  

\bibitem{deHaro:2016ftq} 
  J.~de Haro,
  Gen.\ Rel.\ Grav.\  {\bf 49}, no. 1, 6 (2017)
  [arXiv:1602.07138 [gr-qc]]. 

\bibitem{hap}
J. de Haro, J. Amor\'{o}s and S. Pan, 
Phys. Rev. {\bf D 94}, 064060 (2016) 	[arXiv:1607.06726].


 \bibitem{Geng:2017mic} 
  C.~Q.~Geng, C.~C.~Lee, M.~Sami, E.~N.~Saridakis and A.~A.~Starobinsky,
  JCAP {\bf 1706}, no. 06, 011 (2017)
  [arXiv:1705.01329 [gr-qc]].
  
  
  \bibitem{AresteSalo:2017lkv} 
  L.~Arest\'{e} Sal\'{o} and J.~de Haro,
  Eur.\ Phys.\ J.\ C {\bf 77}, no. 11, 798 (2017)
  [arXiv:1707.02810 [gr-qc]].

\bibitem{Haro:2015ljc} 
  J.~Haro and S.~Pan,
  Int.\ J.\ Mod.\ Phys.\ D {\bf 27}, no. 05, 1850052 (2018)
  [arXiv:1512.03033 [gr-qc]].
	
\bibitem{hyp}
J. Haro, W. Yang and  S. Pan, 	 
JCAP {\bf 01}, 023 (2019)
	[arXiv:1811.07371].	
	
	

\bibitem{Joyce}
M. Joyce, 
Phys. Rev. {\bf D 55}, 1875 (1997)  	[arXiv:9606223].

\bibitem{Parker}
L. Parker, 
Phys. Rev. Lett. {\bf 21}, 562 (1968); L. Parker, 
 Phys. Rev. {\bf 183}, 1057 (1969); L. Parker, 
 Phys. Rev. {\bf D 3}, 346 (1970).


\bibitem{fmm}
V. M. Folov, S. G. Mamayev and V. M. Mostepanenko,
Phys. Lett. {\bf A 55}, 7 (1976).



\bibitem{glm}
A. A. Grib, B. A. Levitskii and V. M. Mostepanenko, 
Theoreticheskaya i Matematicheskaya Fizika {\bf 19}, 59 (1974).


\bibitem{gmm} A. A. Grib, S. G. Mamayev and V. M. Mostepanenko, 
Gen. Rel. Grav. {\bf 7}, 535 (1976); A. A. Grib, S. G. Mamayev and V. M. Mostepanenko, 
Soviet Physics Journal  {\bf 17}, 1700 (1974).


\bibitem{ford}
L. H. Ford, 
Phys. Rev. {\bf D 35}, 2955 (1987).

\bibitem{Zeldovich}
 Ya B. Zeldovich and A. A. Starobinsky, 
 JETP Lett. {\bf 26}, 252 (1977).


\bibitem{Damour} 
T. Damour and A. Vilenkin,
Phys. Rev. {\bf D 53},  2981 (1996) [arXiv:9503149].

\bibitem{Giovannini} M. Giovannini,
 Phys. Rev. {\bf D 58}, 083504 (1998)
[arXiv:9806329].



\bibitem{Spokoiny}
B. Spokoiny, 
Phys. Lett. {\bf B 315}, 40 (1993) [arXiv:9306008].


\bibitem{dimopoulos0} 	
K. Dimopoulos and  C. Owen, 	
 	JCAP {\bf 1706}, 027 (2017)
 	 	[arXiv:1703.00305]


\bibitem{vardayan}
Y. Akrami, R. Kallosh, A. Linde and  V. Vardanyan, 
JCAP {\bf 1806}, 041 (2018) 
 	[arXiv:1712.09693].
	



\bibitem{fkl0}
G. Felder, L. Kofman and  A. Linde, 
 	Phys. Rev. {\bf D 59}, 123523 (1999)  	[arXiv:9812289]




\bibitem{fkl}
G. Felder, L. Kofman and  A. Linde,
Phys. Rev. {\bf D 60}, 103505 (1999) [arXiv:9903350].


\bibitem{dimopoulos} 	
 K. Dimopoulos, L.D. Wood and C. Owen,	
 	  	Phys. Rev. {\bf D 97}, 063525 (2018) 	[arXiv:1712.01760].
 	  	

\bibitem{FL}
B. Feng and  M. Li,
 	Phys. Lett. {\bf B 564}, 169-174 (2003)  	[arXiv:0212213].


\bibitem{ABM}
A. Agarwal, S.  Bekov and  K. Myrzakulov, 
	[arXiv:1807.03629]. 


\bibitem{tommi}
K. Dimopoulos and T. Markkanen,  
JCAP {\bf 06}, 021 (2018)  [arXiv:1803.07399].


\bibitem{kolb}
D. J. H. Chung, E.W. Kolb and A. Riotto,
Phys. Rev.  {\bf D 59}, 023501 (1998) 	[arXiv:9802238].


\bibitem{kolb1}
D. J. H. Chung,  P. Crotty, E. W. Kolb and A. Riotto,
Phys. Rev. {\bf D 64},  043503 (2001) 	[arXiv:0104100].

\bibitem{Birrell1}
  N. D. Birrell and P. C. W. Davies, 
  J. Phys. A: Math. Gen. {\bf 13}, 2109 (1980)


\bibitem{hashiba}
S. Hashiba and J. Yokoyama, 
	[arXiv:1809.05410].

\bibitem{Bunch}
 T. S. Bunch, 
J. Phys.  {\bf A 13}, 1297 (1980).

	


\bibitem{gkr}
G. F. Giudice, E. W. Kolb and  A. Riotto,
 	Phys. Rev. {\bf D 64},   023508 (2001) 	[arXiv:0005123].

\bibitem{lindley}
J. Ellis, D.V. Nanopoulos and  S. Sarkar,
Nuc. Phys. {\bf B 259}, 175 (1985). 


\bibitem{eln}
J. Ellis, A. Linde and D. Nanopoulos, 
Phys. Lett. \textbf{B 118}, 59 (1982).


\bibitem{gmmbook}
A. A. Grib, S.G. Mamayev and V. M. Mostepanenko,
Friedmann Laboratory Publishing for Theoretical Physics,
St. Petersburg (1994).


\bibitem{zs}
Ya B. Zeldodovich  and A. A. Starobinsky,   
Sov. Phys JETP {\bf 34}, 1159 (1972).






\bibitem{bunch}
T. S. Bunch and P. C. W. Davies,
Proc. Roy. Soc. Lond. A {\bf 360}, 177 (1978).


\bibitem{Birrell}
N. D. Birrell  and C. P. W. Davies, {\it  Quantum Fields in Curved Space} (Cambridge: Cambridge University
Press) (1982).

\bibitem{btw}
B. A. Bassett, S. Tsujikawa and D. Wands,
Rev. Mod. Phys. {\bf 78}, 537 (2006) [arXiv:0507632].


\bibitem{hap18}
J. Haro, J. Amor\'{o}s and S. Pan, {\it The Peebles-Vilenkin quintessential inflation model revisited}  (2019)  [arXiv:1901.00167].

{
\bibitem{planck18}
N. Aghanim et al., {\it Planck 2018 results. VI. Cosmological parameters} (2018) [arXiv:1807.06209].

\bibitem{plateau}
  C.~Q.~Geng, C.~C.~Lee, M.~Sami, E.~N.~Saridakis and A.~A.~Starobinsky,
  JCAP {\bf 1706}, no. 06, 011 (2017)  [arXiv:1705.01329].

\bibitem{attractor1}
K. Dimopoulos, L. Donaldson Wood and C. Owen,
Phys. Rev. D {\bf 97}, 063525 (2018) [arXiv:1712.01760].


\bibitem{attractor2}
Y. Akrami, R. Kallosh, A. Linde and V. Vardanyan,
JCAP {\bf1806}, 041 (2018) [arXiv:1712.09693].
}

{

\bibitem{Akrami:2018odb}
  Y.~Akrami {\it et al.} [Planck Collaboration],
  arXiv:1807.06211 [astro-ph.CO].

}

\bibitem{lindebook}
A. Linde,
Contemp. Concepts Phys. {\bf 5},  1 (2005)
	[arXiv:0503203].

\bibitem{ha}
J. de Haro and L. Arest\'e Sal\'o,
Phys. Rev. {\bf D 95}, 123501 (2017) 	[arXiv:1702.04212]. 

\bibitem{Haro} J. Haro, 
J. Phys. A: Mat. Theor. {\bf 44}  205401 (2011).

\bibitem{Winitzki}
S. Winitzki,
 	Phys. Rev. {\bf D 72}, 104011 (2005)  	[arXiv:0510001].


\bibitem{Giovannini99}
M. Giovannini, 
Phys. Rev. {\bf D 60}, 123511 (1999) [arXiv:astro-ph/9903004].


\bibitem{rubio}
J. Rubio and  C. Wetterich, 
 	Phys. Rev. {\bf D 96}, 063509 (2017) 	[arXiv:1705.00552].




{
\bibitem{Enqvist}
K. Enqvist, T. Meriniemi and S. Nurmi,
JCAP{\bf 10}, 057 (2013)  	[arXiv:1306.4511].

\bibitem{Figueroa}
D. G. Figueroa, J. Garcia-Bellido and F. Torrenti,
Phys. Rev. {\bf D92}, 083511 (2015) 	[arXiv:1504.04600].

\bibitem{Freese}
K. Freese, E. I. Sfakianakis, P. Stengel and L. Visinelli,
{\it The Higgs Boson can delay Reheating after Inflation}, (2017)
	[arXiv:1712.03791].
	
	

\bibitem{Turner}
M.S. Turner, Phys. Rev. {\bf 28}, 1243 (1983).	
	
	
	
	}










\end{thebibliography}
\end{document}